\documentclass[twocolumn, 10pt]{IEEEtran}

\usepackage{hyperref} 
\usepackage{graphicx}
\usepackage{amssymb}
\usepackage{amsmath}
\usepackage{mathtools}
\usepackage{cite}
\usepackage{stfloats}
\usepackage{subfigure}
\usepackage{psfrag}
\usepackage[mathscr]{euscript}
\usepackage{acronym}  
\usepackage{booktabs}
\usepackage[table]{xcolor}

\makeatletter
\renewcommand{\fnum@figure}{Fig. \thefigure}
\makeatother

\acrodef{NOMA}{non-orthogonal multiple access}
\acrodef{A2G}{air-to-ground}
\acrodef{A2A}{air-to-air}
\acrodef{ITU}{International Telecommunication Union}
\acrodef{IoT}{Internet of Thing}
\acrodef{D2D}{device-to-device}
\acrodef{BPP}{binomial point process}
\acrodef{MGF}{moment-generating function}
\acrodef{QoS}{quality of service}
\acrodef{QoP}{quality of performance}
\acrodef{GP}{geometric programming}
\acrodef{GGP}{generalized geometric programming}
\acrodef{ASE}{area spectral efficiency}
\acrodef{MISO}{multiple-input single-output}
\acrodef{KKT}{Karush Kuhn Tucker}
\acrodef{MAN}{multilayer aerial network}
\acrodef{AN}{aerial network}
\acrodef{UAV}{unmanned aerial vehicle}
\acrodef{STP}{successful transmission probability}
\acrodef{NLoS}{non-line of sight}
\acrodef{LoS}{line of sight}
\acrodef{MAN}{multi-layer aerial network}
\acrodef{U2U}{\ac{UAV}-to-\ac{UAV}}
\acrodef{U2G}{\ac{UAV}-to-ground}
\acrodef{SINR}{signal to interference plus noise ratio}
\acrodef{PGFL}{probability generating functional}
\acrodef{CCDF}{complementary cumulative distribution function}
\acrodef{CF}{characteristic function}
\acrodef{PPP}{Poisson point process}
\acrodefplural{PPP}{Poisson point processes}
\acrodef{CSI}{channel state information}
\acrodef{OFDM}{orthogonal frequency division multiplexing}
\acrodef{OFDMA}{orthogonal frequency division multiple access}

\acrodef{RV}{random variable}
\acrodef{i.i.d.}{independent, identically distributed}
\acrodef{PMF}{probability mass function}
\acrodef{PDF}{probability distribution function}
\acrodef{CDF}{cumulative distribution function}
\acrodef{ch.f.}{characteristic function}
\acrodef{AWGN}{additive white Gaussian noise}
\acrodef{SNR}{signal-to-noise ratio}
\acrodef{LRT}{likelihood ratio test}
\acrodef{DRT}{distance ratio test}
\acrodef{GLRT}{generalized likelihood ratio test}
\acrodef{CRLB}{Cram\'{e}r-Rao lower bound}
\acrodef{CRB}{Cram\'{e}r-Rao bound}
\acrodef{ZZLB}{Ziv-Zakai lower bound}
\acrodef{ZZB}{Ziv-Zakai bound}
\acrodef{LOS}{line-of-sight}
\acrodef{ToF}{time-of-flight}
\acrodef{NLOS}{non-line-of-sight}
\acrodef{GDOP}{geometric dilution of precision}
\acrodef{GPS}{Global Positioning System}
\acrodef{FIM}{Fisher information matrix}
\acrodef{PEB}{position error bound}
\acrodef{SPEB}{squared position error bound}
\acrodef{TOA}{time-of-arrival}
\acrodef{TOF}{time-of-flight}
\acrodef{WSN}{wireless sensor network}
\acrodef{MAC}{medium access control}
\acrodef{RSS}{received signal strength}
\acrodef{WAF}{wall attenuation factor}
\acrodef{TDOA}{time difference-of-arrival}
\acrodef{RF}{radiofrequency}
\acrodef{RTT}{round-trip time}
\acrodef{AOA}{angle-of-arrival}
\acrodef{MF}{matched filter}
\acrodef{ED}{energy detector}
\acrodef{ML}{maximum likelihood}
\acrodef{MSE}{mean-square error}
\acrodef{RMSE}{root-mean-square error}
\acrodef{LEO}{localization error outage}
\acrodef{ppm}{part-per-million}
\acrodef{ACK}{acknowledge}
\acrodef{UWB}{Ultrawide bandwidth}
\acrodef{TNR}{threshold-to-noise ratio}
\acrodef{LS}{least squares}
\acrodef{IR-UWB}{impulse radio UWB}
\acrodef{FCC}{Federal Communications Commission}
\acrodef{TH}{time-hopping}
\acrodef{PPM}{pulse position modulation}
\acrodef{MUI}{multi-user interference}
\acrodef{PDP}{power delay profile}
\acrodef{BPZF}{band-pass zonal filter}
\acrodef{SIR}{signal-to-interference ratio}
\acrodef{RFID}{radio frequency identification}
\acrodef{WPAN}{wireless personal area network}
\acrodef{WWB}{Weiss-Weinstein bound}
\acrodef{DP}{direct path}
\acrodef{MF}{matched filter}
\acrodef{MMSE}{minimum-mean-square-error}
\acrodef{SBS}{serial backward search}
\acrodef{SBSMC}{serial backward search for multiple clusters}
\acrodef{NBI}{narrowband interference}
\acrodef{WBI}{wideband interference}
\acrodef{INR}{interference-to-noise ratio}
\acrodef{CR}{channel response}
\acrodef{CIR}{channel impulse response}
\acrodef{CR}{channel  response}
\acrodef{RADAR}{radar}
\acrodef{MUR}{Multistatic radar}
\acrodef{JBSF}{jump back and search forward}
\acrodef{HDSA}{high-definition situation-aware}
\acrodef{RRC}{root raised cosine}
\acrodef{ST}{simple thresholding}
\acrodef{BTB}{Bellini-Tartara bound}
\acrodef{P-Max}{$P$-Max}  
\acrodef{MIMO}{multiple-input multiple-output}
\acrodef{MAP}{maximum a posteriori}
\acrodef{FG}{factor graph}
\acrodef{OP}{outage probability}
\acrodef{WED}{wall extra delay}
\acrodef{RMS}{root mean square}
\acrodef{SPAWN}{sum-product algorithm over a wireless network}
\acrodef{MDD}{minimum distance distribution}
\acrodef{MAP}{maximum a posteriori probability}
\acrodef{PAR}{probabilistic association rule}

\usepackage{color}
\usepackage{dsfont}
\usepackage{bbm}

\newcommand{\RTsel}{a receiver in the $i$-layer and a transmitter in the $j$-layer under the channel environment $c$}

\newcommand{\oT}{\text{t}}
\newcommand{\oR}{\text{r}}
\newcommand{\aN}{\text{n}}
\newcommand{\aS}{\text{s}}
\newcommand{\Rx}{\text{Rx}}
\newcommand{\Tx}{\text{Tx}}

\newcommand{\ijtau}{_{ij, \tau}}
\newcommand{\ijo}{_{ij, o}}
\newcommand{\ko}{_{k, o}}
\newcommand{\kr}{_{k, \oR}}
\newcommand{\kt}{_{k, \oT}}

\newcommand{\krx}{_{k,\Rx}}
\newcommand{\ktx}{_{k,\Tx}}

\newcommand{\ktxn}{_{k}}
\newcommand{\jtx}{_{j, \Tx}}
\newcommand{\irx}{_{i, \Rx}}
\newcommand{\jtxn}{_{j}}

\newcommand{\ijtx}{_{ij, \Tx}}
\newcommand{\ijrx}{_{ij, \Rx}}
\newcommand{\iktx}{_{ik, \Tx}}

\newcommand{\ctau}{\chi\ijk{\tau}(y)}

\def\cb{c_\text{o}}
\newcommand{\STP}[1]{\mathcal{P}_{#1,\tau}}
\newcommand{\ASE}[1]{\mathcal{S}_{#1,\tau}}

\newcommand{\setTx}{\mathcal{T}}

\newcommand{\Lb}[1]{\lambda_{j, \Tx}^{b,#1}}

\newcommand{\oa}{\tau}
\newcommand{\MinP}{\mainA\in\Phi\ijo\Co}
\newcommand{\mainA}{\bx_\tau}

\newcommand{\Yoa}{Y_{ij,\oa}\Co}
\newcommand{\Vj}{V_{ij,o}\Co}
\newcommand{\Vk}{V_{ik,o}\To}
\newcommand{\bxInt}{_{\bx\in\Phi_{ik,\text{Tx}}\To/\mainA }}
\newcommand{\Emain}{\varepsilon\ijtau\Co(y)}
\newcommand{\fS}{f_{\mathcal{P}}(i,j,c,\tau)}

\newcommand{\frS}{f_{\mathcal{P}}(k,j,c,\tau)}
\newcommand{\ftS}{f_{\mathcal{P}}(i,k,c,\tau)}

\newcommand{\AssoM}{\mathcal{A}_{ij,\tau}\Co}
\newcommand{\LapS}{l_{ij}\Co(y)}
\newcommand{\poa}[1]{p_{ij,#1}\Co}
\newcommand{\PIjk}{\prod_{\substack{k\in\mathcal{K}, \Tin,\\(k, \cb)\neq(j,c)}}}

\newcommand{\Rjk}{R\ijk{a}}
\newcommand{\ijk}[1]{_{j,k,#1}\CTo}
\newcommand{\hij}{h_{ij}}
\newcommand{\hik}{h_{ik}}
\newcommand{\hkj}{h_{kj}}

\newcommand{\sumAp}{\sum_{	i \in \mathcal{K},\Cin}  \int_{h_{ki}}^\infty}
\newcommand{\sumApj}{\sum_{	i \in \mathcal{K},\Cin}  \int_{h_{ik}}^\infty}
\newcommand{\fSi}{\mathcal{C}_{ki,\tau}\Co(y)}
\newcommand{\fSj}{\mathcal{C}_{ik,\tau}\Co(y)}
\newcommand{\pkj}{\phi_{kj, \tau}\Co(i, y)}
\newcommand{\pkjj}{\phi_{kj, \tau}\Co(j, y)}
\newcommand{\thkj}{\theta_{kj, \tau}\Co(i, y)}

\newcommand{\bx}{\mathbf{x}}
\newcommand{\ijCo}{_{ij}\Co}

\newcommand{\ikTo}{_{ik}\To}

\newcommand{\ijcoL}{_{ij}\coL}
\newcommand{\ijcoN}{_{ij}\coN}

\newcommand{\Co}{^{(c)}}
\newcommand{\To}{^{(\cb)}}
\newcommand{\CTo}{^{(c, \cb)}}
\newcommand{\coL}{^{\text{(L)}}}
\newcommand{\coN}{^{\text{(N)}}}

\newcommand{\aCo}{^{-\alpha\Co}}
\newcommand{\aTo}{^{-\alpha\To}}
\newcommand{\acoN}{^{-\alpha\coN}}
\newcommand{\acoL}{^{-\alpha\coL}}

\newcommand{\Lap}[1]{\mathcal{L}_{#1}}

\newcommand{\Cin}{c\in\text{\{L,N\}}}
\newcommand{\Tin}{\cb\in\text{\{L,N\}}}

\newtheorem{lemma}{Lemma}
\newtheorem{corollary}{Corollary}

\newtheorem{remark}{Remark}

\acrodef{BS}{base station}
\acrodef{AP}{access point}

\begin{document}

	\newcommand{\paperTitle}{
		Multi-layer Unmanned Aerial Vehicle Networks:\\Modeling and Performance Analysis
	}

	\title{\paperTitle}
	\author{
		\vspace{0.2cm}
		Dongsun~Kim,
		Jemin~Lee,\textit{ Member, IEEE},
		and Tony Q. S. Quek,\textit{ Fellow, IEEE}
		%
		%
		%
		\thanks{
D.\ Kim and J.\ Lee are with the Department of Information and Communication Engineering, Daegu Gyeongbuk Institute of Science and Technology, Daegu 42988, South Korea (e-mail: yidaever@dgist.ac.kr; jmnlee@dgist.ac.kr).
		}
	\thanks{T.\ Q.\ S.\ Quek is with Information Systems Technology and Design Pillar, Singapore University of Technology and Design, Singapore 487372 (e-mail: tonyquek@sutd.edu.sg).}
		\thanks{
			Part of this paper was presented
			at the IEEE Global Communications Conference, UAE, December 2018 \cite{DonJemTon:18}.
		}
	}
	\maketitle 
	\setcounter{page}{1}
	\acresetall
	\begin{abstract}



Since various types of unmanned aerial vehicles (UAVs) with different hardware capabilities are introduced, we establish a foundation for the multi-layer aerial network (MAN). First, the MAN is modeled as $K$ layer ANs, and each layer has UAVs with different densities, floating altitudes, and transmission power. To make the framework applicable for various scenarios in MAN, we consider the transmitter- and the receiver-oriented node association rules as well as the air-to-ground and air-to-air channel models, which form line of sight links with a location-dependent probability. We then newly analyze the association probability, the main link distance distribution, successful transmission probability (STP), and area spectral efficiency (ASE) of MAN. The upper bounds of the optimal densities that maximize STP and ASE are also provided. Finally, in the numerical results, we show the optimal UAV densities of an AN that maximize the ASE and the STP decrease with the altitude of the network. We also show that when the total UAV density is fixed for two layer AN, the use of single layer in higher(lower) altitude only for all UAVs can achieve better performance for low(high) total density case, otherwise, distributing UAVs in two layers, i.e., MAN, achieves better performance.

	\end{abstract}

	\acresetall


	\section{Introduction}

	Recent development of the \ac{UAV} technologies enables 
the \ac{UAV} to play various roles in the wireless networks. 
The \acp{UAV} are expected to work as temporal base stations 
in case of the disaster and the data demanding events\cite{ZenZhaLim:16},
and the data acquisition for the crowd surveillance can also be done by \acp{UAV}\cite{MotBagTal:17}.
Furthermore, the \acp{UAV} can act as a relay for unreliable direct link case\cite{ZengZhaLim:16}.
As such demands on the \ac{UAV} communications and the number of \acp{UAV} increase, 
the research for the reliable \ac{AN} must be preceded.

The \ac{UAV} based wireless communication has been studied in \cite{ AlKanLar:14, AlKanJam:14,YanZhoZha:18 ,MozSaaBen:16, CheTonJem:17, MinJem:19} after modeling the wireless channel and the mobility, which are different from those of the terrestrial networks.
In \cite{AlKanLar:14}, the probability that a link forms \ac{LoS}, i.e., the \ac{LoS} probability, is modeled, which is determined by the angle from the ground, and also proposed the optimal \ac{UAV} deployment that maximizes the coverage area.
The pathloss and the channel gain of the link between a \ac{UAV} and a ground node are studied in \cite{AlKanJam:14}.
In \cite{YanZhoZha:18}, the \ac{LoS} probability is provided for the link between \acp{UAV}, which have different altitudes.
Considering LoS channel, device-to-device communications, secrecy capacity \ac{UAV}-aided communication systems, and \ac{UAV} to ground communication in presence of interferer are studied in \cite{MozSaaBen:16}, \cite{CheTonJem:17}, and \cite{MinJem:19}, respectively.
However, the studies mentioned above have considered only the small number of \acp{UAV}, which show the performance of the limited \ac{UAV} communication scenarios.

Recently, the researches on the \acp{AN}, which is the wireless networks consisting of multiple \acp{UAV}, have been presented in \cite{ MohYurOsa:17, MohHal:18, HouLiuZhe:18, JinYou:17,SekTabHos:18, ZhaZha:16, QiYanPen:18, HuiXiaNin:18, AzaRosChi:17, RabLutHes:18, AliSyeDes:18, EsmMce:18}.
For those works, the stochastic geometry, which is a widely-used tool for randomly distributed nodes\cite{HaeGan:09}, has been used.
The \ac{PPP}-based \acp{AN} model is presented and studied in \cite{ MohYurOsa:17, MohHal:18, HouLiuZhe:18}, and the \ac{LoS} and \ac{NLoS} channels are considered for the \ac{A2G} communications in \cite{ MohYurOsa:17} and \cite{MohHal:18}. 
Furthermore, the research on the coexistence of \acp{AN} with the terrestrial network is presented in \cite{ ZhaZha:16, QiYanPen:18, HuiXiaNin:18, AzaRosChi:17, RabLutHes:18, AliSyeDes:18, EsmMce:18}.
In these works, the terrestrial network is modeled by a \ac{PPP} and
the distribution of \acp{UAV} is modeled by 3-D \ac{PPP}\cite{ZhaZha:16, QiYanPen:18, HuiXiaNin:18} and 2-D \ac{PPP}\cite{AzaRosChi:17, RabLutHes:18}. 
Especially, in \cite{AliSyeDes:18} and \cite{EsmMce:18}, 
the random distribution of users are also considered and modeled by a clustering point process (for disaster area or temporal data demanding events like concert)\cite{AliSyeDes:18} and a \ac{PPP}\cite{EsmMce:18}.
However, most of these works did not consider the multiple layer structure of \ac{AN}, of each layer has different types of \acp{UAV}.


The \acp{AN} can have various types of \acp{UAV} with different floating altitudes and transmission power depending on their hardware constraints\cite{ChaGomAl:16}, 
which leads to the \emph{multiple layer structure} in \ac{AN}. 
The multiple layer structure can also be useful and required
for better resource management and reliable communications, 
especially when the number of \acp{UAV} and \ac{UAV}-related applications increase.  
Recently, the multiple layer structure for \ac{UAV} communications has been considered in \cite{JinYou:17}, \cite{SekTabHos:18}, and \cite{EsmMce:18}. 
In \cite{JinYou:17} and \cite{SekTabHos:18}, 
the \acp{UAV} are used as relays \cite{JinYou:17} or downlink base stations to improve the downlink spectral efficiency \cite{SekTabHos:18}. 
However, a analysis result on the performance has not been provided, 
especially in terms of the \ac{STP} or the \ac{ASE} of the multiple layer structure \ac{AN}.
In \cite{EsmMce:18}, the spectral efficiency of multiple layer structure \ac{AN} was analyzed, by focusing on the communications of ground base stations, which are assisted by \acp{UAV}.
However, in \cite{EsmMce:18}, only the performance of the single layer \ac{AN} case is provided in the simulation results and the communication between \acp{UAV} are not considered, which fails to fully explore the efficient design of the multiple layer \ac{AN}. 


%
%

Therefore, in this paper, we consider the \ac{MAN}, and provide a framework for the efficient design of the \ac{MAN}. 
We first model the \ac{MAN}, which is composed of $K$ layers of \acp{AN} including \acp{UAV} with different transmission power, spatial densities, and floating altitudes.
We then analyze the \ac{STP} and the \ac{ASE} of the \ac{MAN}, and explore how to design the \ac{MAN} to maximize its performance. 
The contribution of this work can be summarized as follows.

\begin{itemize}
\item Differently to prior works on aerial networks and terrestrial heterogeneous networks, 
	we model the \ac{MAN} by considering both the node association rules and channel model, suitable for various scenarios of the \ac{MAN}. 
	Specifically, 
according to the association subject, we consider two types: 
the \emph{transmitter-oriented} association (e.g., when a transmitting \ac{UAV} selects the best receiving \ac{BS}) 
and the \emph{receiver-oriented} association (e.g., when a receiving \ac{UAV} selects the best transmitting \ac{BS}).
Furthermore, we consider both \ac{A2A} and \ac{A2G} channels, 
which form \ac{LoS} links with a certain probability, determined by not only the link distance but also the \ac{UAV} altitude.
\item We newly analyze the Laplace transform of the interference considering LoS and NLoS channels with the LoS probability, and 
	provide that of the interference from same layer UAVs in a closed form. 
	Note that the multiple layer structure has been considered for terrestrial networks, 
	called as the heterogeneous networks \cite{DhiGanBacAnd:12, SinDhiAnd:13, ZhaYanQueLee:17},
	and the Laplace transform of the interference has also been analyzed. 
	However, as the node association rules and the channel model, suitable for \ac{MAN}, are used in this work, 
	the analysis has been newly performed.
\item We then analyze the \ac{STP} and the \ac{ASE} of the \ac{MAN} using stochastic geometry. 
	We also provide the upper bound of the optimal transmitting UAV densities for each layers, which maximize the \ac{STP} and the \ac{ASE}.
	This is the first work, optimizing the node density of \ac{AN}, to the best knowledge of the authors.
	%
	%
%
\item We finally provide insights on the efficient design of \ac{MAN} via numerical results. 
	Specifically, we provide the optimal altitude and the densities of UAV in each layer in terms of the \ac{STP} and the \ac{ASE},
	and also show when the multiple layer structure of \ac{AN} can achieve better performance than the single layer \ac{AN}.
\end{itemize}

\begin{table}
\caption{Notations used throughout the paper.} \label{table:notation}
\begin{center}
\rowcolors{2}
{cyan!15!}{}
\renewcommand{\arraystretch}{1.2}
\begin{tabular}{c  p{6cm} }
\hline
 {\bf Notation} & {\hspace{2.5cm}}{\bf Definition}
\\
\midrule
\hline
$\mathcal{K}$ & Set of layers constituting the \ac{MAN} \\ \addlinespace
$h_k $ & altitude of the $k$-layer nodes \\ \addlinespace
$P_k $ & Transmission power of the $k$-lyaer nodes \\ \addlinespace
$\lambda_{k, \Rx(\Tx)}$ & Density of the receiver(transmitter) in the $k$-layer \\ \addlinespace
$\Phi_{k, \Rx(\Tx)}$ & Distribution of the receiver(transmitter) in the $k$-layer \\ \addlinespace
$\Cin$& Indicator whether the channel is \ac{LoS} or \ac{NLoS}\\ \addlinespace
$\alpha\Co$ & Pathloss exponent of channel $c$ \\ \addlinespace
$G^{m\Co}$ & Channel gain of channel $c$ \\ \addlinespace
$\rho\ijCo (x)$ & Probability that link between the $i$-layer receiver and the $j$-layer transmitter is under channel environment $c$ when the link distance is $x$ \\ \addlinespace
$\tau=oa $ & Communication node association rule defined by $o$ and $a$ \\ \addlinespace
$o\in\{\oR,\oT\} $ & Node association rule that indicates whether the communication is the receiver-oriented or the transmission oriented association \\ \addlinespace
$a\in\{\aN,\aS\} $ & Node association rule that indicates whether the node with the the nearest distance or the strongest power is selected \\ \addlinespace
$\AssoM $ & Probability that the main link is established between the $i$ and $j$-layer nodes under channel $c$ using association rule $\tau$ \\ \addlinespace
$\Yoa$ & Random variable that represents the main link distance given association $\AssoM$ \\ \addlinespace
$I_{ij}\Co$ & Interference to the $i$-layer receiver from $j$-layer transmitters in the channel $c$ \\ \addlinespace
$\mathcal{I}_i$ & Sum of the interference and noise to the $i$-layer receiver \\ \addlinespace
$\chi$ & Distance that indicates the area where the interferer cannot exist\\ \addlinespace
$\Emain$ & Event when the main link is established between $i$ and $j$-layer with distance $y$ using the node association rule $\tau$\\ \addlinespace
$\mathcal{L}_{I|\Emain} $ &Laplace transform of the $I$ in the event of $\Emain$
\\ \addlinespace
$p_{ij,\tau}\Co(y)$ & \ac{STP} in the event of $\Emain$\\ \addlinespace
$\STP{k} $ & \ac{STP} of the $k$-layer in the \ac{MAN} \\ \addlinespace
$\ASE{k} $ & \ac{STP} of the $k$-layer in the \ac{MAN} [bps/Hz/$\text{m}^2$] \\ \addlinespace
$\Lb{\STP{k}(\ASE{k})} $ & Upper bound of the optimal transmitter density in the $j$-layer that maximizes the \ac{STP}(\ac{ASE}) of $k$-layer\\ \addlinespace


\hline
\end{tabular}
\end{center}\vspace{-0.63cm}
\end{table}%

	\section{System Model}
	In this section, we present the system model of a \ac{MAN} including the network description and the channel model. Furthermore, we describe the node association rules and present the \ac{PDF} of the main link distance.

	\subsection{Multi-layer Aerial Network Structure}

	\begin{figure}[t!]
		\begin{center}
			{
				\includegraphics[width=1.00\columnwidth]{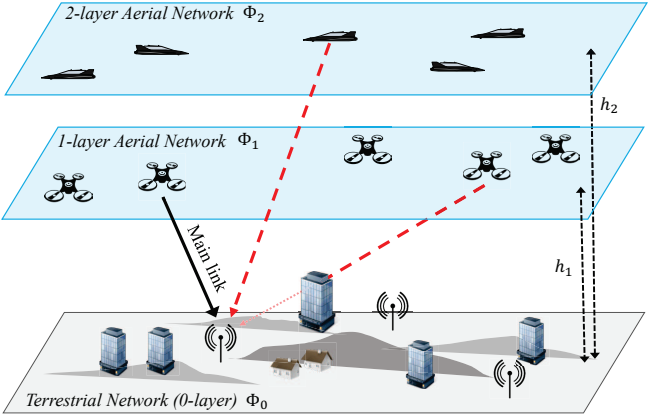}
			}
		\end{center}
		\caption{
			An example of two layer \ac{MAN} with ground receivers (i.e., $0$-layer).
			The black lines represent the main link and red dashed lines represent interference links.
		}   \label{fig:System}
	\end{figure}

We consider a \ac{MAN}, which consists of $K$ layers of \ac{AN} at different altitudes as shown in Fig.~\ref{fig:System}. We denote $\mathcal{K}$ as the set of layers constituting the \ac{MAN}, i.e., $\mathcal{K}=\{0, 1, \cdots, K\}$, where layer $0$ is the terrestrial network. We assume \acp{UAV} in \acp{AN} are distributed according to \acp{PPP} such as in \cite{AzaRosChi:17,GalKibDas:17} as well as the ground nodes in the terrestrial network \cite{HaeGan:09}. 
Specifically, in the $k$-layer, the node locations follow a homogeneous \ac{PPP} $\Phi_k$ with density $\lambda_k$, and they are at the fixed altitude $h_k$ and transmit with the power $P_k$.
In the $k$-layer, nodes act as either a receiver or a transmitter, where the set of the receivers and the transmitters are denoted by $\Phi\krx$ and $\Phi\ktx$. 
Similarly, the densities of the receivers and the transmitters in the $k$-layer are given by $\lambda\krx$ and $\lambda\ktx$. 
Here, $\Phi_k=\Phi\krx+\Phi\ktx$ and $\lambda_k=\lambda\krx+\lambda\ktx$.
The altitude of nodes in the $0$-layer (i.e., the terrestrial layer) is $h_0 = 0$ and altitudes of other layers are $h_k \geq0$ for $k \in \mathcal{K}$.
In addition, the altitude between the $i$-layer and the $j$-layer is denoted by $\hij=|h_i-h_j|$. 
%

\subsection{Channel Model}
In the terrestrial network, where the transmitter and the receiver are on the ground, the channel is generally modeled as \ac{NLoS} links 
However, in the \ac{MAN}, we have the communication between a \ac{UAV} and a ground node and the communication between \acp{UAV}.
For those communications, we consider both \ac{LoS} and \ac{NLoS} links, which are affected by the existence of obstacles (e.g., buildings) between the transmitter and the receiver \cite{AlKanLar:14, YanZhoZha:18} by following the ITU model \cite{ITUPDP:15}.
In this paper, we define the probability of forming \ac{LoS} channel as the \ac{LoS} probability $\rho\ijcoL(x)$ and the probability of forming \ac{NLoS} channel as the \ac{NLoS} probability $\rho\ijcoN(x)=1-\rho\ijcoL(x)$, where a receiver and a transmitter are in the $i$-layer and the $j$-layer, respectively, and the link distance is $x$.
From \cite{AlKanLar:14}, the \ac{LoS} probability is given by
\begin{align}\label{eq:LoS_P_origin}
\rho\ijcoL(x)=\!\prod_{n=0}^{m}\left[ 1-\exp\left(	-\frac{\left[\max(h_i, h_j)-\frac{(n+1/2)\hij}{ m+1}\right]^2}{2\xi^2}	\right) \right]
\end{align}
where $m= \text{floor}\left(\sqrt{(x^2-\hij^2)\mu\nu}-1\right)$. 
Here, $\mu$, $\nu$, and $\xi$ are the parameters related to the environments \cite{ITUPDP:15}.
Specifically, $\mu$ is the ratio of area covered by buildings to total area, $\nu$ is the mean number of buildings per unit area, and $\xi$ is the average altitude of the buildings.
%
%
The \ac{LoS} probability can also be approximately determined using the sigmoid function based approximation \cite{AlKanLar:14, BaiVazHea:14}.\footnote{Similar result with the same approach is provided in the \cite{YanZhoZha:18}, however, we follow the \cite{BaiVazHea:14} to provide well-matched approximation with our model.}
Specifically, for the \ac{A2G} channel (i.e., $i$ or $j$=0, $i \neq j$), $\rho\ijcoL(x)$ is given by \cite{AlKanLar:14}
\begin{align}\label{eq:LoS_A2G}
\rho\ijcoL(x)\approx
\frac{1}{1+\iota\exp(-\kappa\left[\sin^{-1}\left(\frac{\hij}{x}\right)-\iota\right])}
\end{align}
where $\iota$ and $\kappa$ are related to $\nu$, $\mu$, and $\xi$ \cite{AlKanLar:14}.
For the \ac{A2A} channel (i.e., $i$ and $j \neq 0$), using the exponential function based approximation $\rho\ijcoL(x)$ is given by \cite{BaiVazHea:14} 
\begin{align}\label{eq:LoS_A2A}
&\rho\ijcoL(x)\approx  \\
&\begin{cases}
\!\left(1-\exp\left\{-\frac{h_i^2}{2\xi^2}\right\}\right)^{ x \sqrt{\nu\mu}} &\text{for } i = j,  \\ 
\!\left(1-\frac{\sqrt{2\pi}\xi}{\hij}\left|Q\left(\frac{h_i}{\xi}\right)-Q\left(\frac{h_j}{\xi}\right)\right|\right)^{ \sqrt{(x^2-\hij^2)\nu\mu}} &\text{for } i\neq j, 
\end{cases}\nonumber
\end{align}
where $Q(x)=\int_x^\infty \frac{1}{\sqrt{2\pi}}\exp\left(-\frac{x^2}{2}\right)dx$.
Especially, the \ac{LoS} channel between ground nodes is given by $\rho\ijcoL(x)=0$.

\begin{figure}[t!]
	\begin{center}
		{
			\psfrag{AAAAAAAAAAAAAAAAAAAAA}[][][0.9]{$h_i=0,\; h_j=40$ (A2G)}
			\psfrag{AAAAAAAAAAAAAAAAAAAAB}[][][0.9]{$h_i=20, h_j=40$ (A2A)}
			\psfrag{AAAAAAAAAAAAAAAAAAAAC}[][][0.9]{$h_i=20, h_j=60$ (A2A)}
			\psfrag{AAAAAAAAAAAAAAAAAAAAD}[][][0.9]{$h_i=40, h_j=40$ (A2A)}
			\psfrag{EAAAAAAAAAA}[][][0.8]{Horizontal link distance, $r$ (m)}
			\psfrag{FAAAAAAAAAA}[][][0.8]{\ac{LoS} Probability}
			\includegraphics[width=1.00\columnwidth]{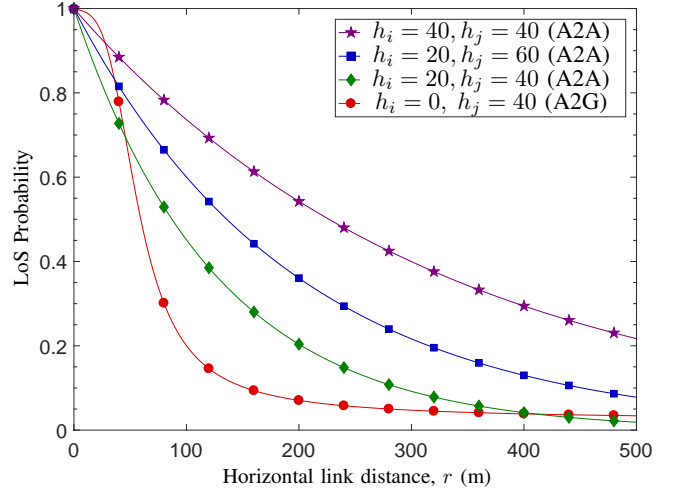}
			\vspace{-8mm}
		}
	\end{center}
\caption{
	\ac{LoS} channel probabilty according to the horizontal distance $r$ with different receiver and transmitter altitudes $h_i$ and $h_j$.
}   \label{fig:LoSP}
\end{figure}

From \eqref{eq:LoS_A2G} and \eqref{eq:LoS_A2A}, we can see that the \ac{LoS} probability is affected by both the horizontal distance and the altitude difference between the transmitter and the receiver, and this is also shown in Fig.~\ref{fig:LoSP}. 
	%
	Figure \ref{fig:LoSP} show  the \ac{LoS} probability as a function of the horizontal distance $r=\sqrt{x^2-\hij^2}$ for \ac{A2A} and \ac{A2G} channels in the dense urban environment.\footnote{The parameters used in this figure are $\mu=0.5$, $\nu=3\times 10^{-4}$ (buildings/$\text{m}^2$), $\xi=20$ (m), $a=12.0810$, and $b=0.1139$  \cite{AlKanLar:14}.}
We observe that the \ac{LoS} probability decreases with $r$ since the larger distance generally has more blockages, which causes the \ac{NLoS} environment. On the contrary, the \ac{LoS} probability increases not only with the altitude difference between the transmitter and receiver, $\hij$, but also with altitudes of the transmitter and the receiver, $h_i$ and $h_j$.

Based on the \ac{LoS} probability, we define $\Phi\ijtx\coL$ and $\Phi\ijtx\coN$ as the set of transmitters in the $j$-layer which have \ac{LoS} and \ac{NLoS} channels to a receiver in the $i$-layer.
Similarly, we define $\Phi\ijrx\coL$ and $\Phi\ijrx\coN$ as the set of receivers in the $i$-layer, which have \ac{LoS} and \ac{NLoS} channels to a transmitter in the $j$-layer, respectively.
Here, the density of $\Phi\ijtx\Co$ and $\Phi\ijrx\Co$ for given link distance $x$ becomes $2\pi x\lambda\jtx\rho\ijCo (x)$ and $2\pi x\lambda\irx\rho\ijCo (x)$, $\forall c=\{\text{L, N}\}$, respectively.



The pathloss exponents for \ac{LoS} and the \ac{NLoS} channels are denoted by $\alpha\coL$ and $\alpha\coN$, respectively, and generally, $2\leq \alpha^{(\text{L})} \leq\alpha^{(\text{N})}$. 
We also consider the Nakagami-$m$ fading for \ac{LoS} and the \ac{NLoS} channels,
of which channel gains are respectively presented by
$G\coL \sim \Gamma(m\coL,\frac{1}{m\coL})$ and $G\coN \sim \Gamma(m\coN,\frac{1}{m\coN})$.
Here, we use $m\coN=1$, which gives Rayleigh fading, i.e., $G\coN\sim\exp(1)$,
while $m\coL \geq 1$.

%

	\subsection{Communication Node Association Rules}\label{SS:AssoRules}
	For node association rules, we consider two components $o$ and $a$: 
	\begin{itemize}
		\item[1)] association subject $o$ (whether a transmitter/receiver selects a receiver/transmitter); and 
		\item[2)] association criterion $a$ (whether the node with strongest power or the nearest distance is selected).
\end{itemize}
%
%

	In a \ac{MAN}, a \ac{UAV} may need to receive data from a ground base station and a \ac{UAV}.
	For those cases, the \ac{UAV} (i.e., receiver) can select the best transmitter for reliable communication and we call it as the \textit{receiver-oriented} association, denoted by $o=\oR$.
	On the other hand, a \ac{UAV} may need to transmit to a ground base station or a \ac{UAV}.
	For those cases, the \ac{UAV} (i.e., transmitter) can select the best receiver, and we call it as the \textit{transmitter-oriented} association, denoted by $o=\oT$.
	For the selection criterion$a$, we consider both the nearest distance association and the strongest power association, denoted by as $a=\aN$ and $a=\aS$, respectively. 
	Note that the nearest and the strongest power associations have been generally used in wireless networks \cite{ChoLiuLee:18}.

	Based on the node association rule $\tau=oa$, the coordinate of the associated node for a node at $\bx$ is defined as
	\begin{align}\label{eq:tau_rule}
	\mainA	=
	\begin{cases}
	\underset{\bx\in\Phi\ko, k\in\mathcal{K}}{\operatorname{arg\,max}} B_k | \bx-\hat{\bx}|^{-1} & \text{for}\;a=\aN,   \\
	\underset{\bx\in\Phi\ko, k\in\mathcal{K}}{\operatorname{arg\,max}} B_k | \bx-\hat{\bx} |^{-\alpha_\bx} & \text{for}\;a=\aS ,
	\end{cases}
	\end{align}
	%
	%
	where $B_k$ is the association bias of $k$-layer, $\alpha_\bx$ is the pathloss exponent of the link between a transmitter and a receiver where the node $\bx$ is involved.
	In \eqref{eq:tau_rule}, $\Phi\ko$ is defined as $\Phi\kr=\Phi\ktx$ and $\Phi\kt=\Phi\krx$.
%

\subsection{Main Link Distance Distribution Analysis}

In the conventional terrestrial networks, the \ac{PDF} of the main link distance is determined by the transmission power, the pathloss exponent, and the link distance.
However, in the \ac{AN}, we need to consider the \ac{LoS}/\ac{NLoS} probabilities for the links. 
Using the association $\tau=oa$, the \ac{PDF} of the main link distance is presented in the following lemma.
We use $f_X(x)$, $F_X(x)$, and $\bar{F}_X(x)$ to represent the \ac{PDF}, \ac{CDF}, and \ac{CCDF} of a random variable $X$, respectively. 

\begin{lemma}\label{L:PDFM}
	Using the node association rule $\tau$, when main link is established between \RTsel,
	the \ac{PDF} of main link distance $\Yoa$ is given by
	\begin{align}
	&f_{\Yoa} (y)
	=
	\frac{f_{\Vj}(y)}{\AssoM}
	\PIjk
	\bar{F}_{\Vk}\left(  \Rjk (y)\right),\label{eq:tau_PDF}
	\end{align}
	where $\AssoM$ is the association probability given by
	\begin{align}
	\AssoM	&= \int\displaylimits_{x>0}
	f_{\Vj} (x)	\PIjk \bar{F}_{\Vk}\left(  \Rjk(x) \right) dx. \label{eq:tau_Asso}
	\end{align}
	Here, $\Vk$ is the distance to the nearest node among the nodes in the $k$-layer \ac{AN} under the channel environment $\Tin$ from a node in the $i$-layer,
	of which the \ac{CCDF} and the \ac{PDF} are given by
	\begin{align}\label{eq:ND_PDF}
	&f_{\Vk}(v)=2\pi\lambda\ko v \rho\ikTo\! (v)\!\exp\left\{\!-\! \int_{\hik}^{v}\! 2\pi\lambda\ko x \rho\ikTo\! (x)dx \!\right\}\!, \nonumber \\
	&\bar{F}_{\Vk}(v)=\exp\left[- \int_{\hik}^{\max(v,\hik)} 2\pi\lambda\ko x \rho\ikTo  (x)dx \right],
	\end{align}
	{where $f_{\Vk}(v)=0\text{ if }v\geq\hik$.} 
	In addition, $\Rjk(y)$ is given by
	\begin{align}\label{eq:int_dis}
	\Rjk(y)=
	\begin{cases}
	y B_k/B_j & \text{for}\;a=\text{n},   \\
	\left(y^{\alpha^{(c)}}B_k/B_j\right)^{1/\alpha^{(\cb)}} & \text{for}\;a=\text{s}.  \\
	\end{cases}
	\end{align}
\end{lemma}

	\begin{IEEEproof}
		From the \ac{LoS} probability, the density in the $k$-layer \ac{AN} under the channel environment $\cb$ in distance $x$ is given by $2\pi x\lambda\ko \rho\ikTo (x)$. Therefore, the \ac{CDF} of $\Vk$ is given by
		\begin{align}
		F_{\Vk}(v)	&
		\overset{(a)}{=}
		1-\exp\left\{-\int_{\hik}^{\max (v, \hik)} 2\pi x\lambda\ko \rho\ikTo (x) dx\right\}\label{eq:ND_CDF}
		\end{align}
where (a) is from the void probability of \ac{PPP}. From \eqref{eq:ND_CDF}, we have \eqref{eq:ND_PDF}.

In the nearest distance association case, the main link has the smallest distance, hence, the probability that main link is established as $\MinP$ and the main link distance is smaller than $y$ is given by
\begin{align}
&\mathbb{P}\left(\Vj\leq y ,\; \MinP \;|\; a=\aN \right) \label{eq:Nn_proof}  \\
&=\int_{0}^{y} f_{\Vj}(v) \mathbb{P}\left(\MinP \;\biggr|\; \Vj=v, a=\aN \right) d v  \nonumber \\
&\overset{(a)}{=}\int_0^{y} f_{\Vj}(v) \!\!\! \PIjk \!\!\! \mathbb{P}\left[  \frac{B_j}{v} \geq \frac{B_k}{\Vk}    \right] dv \nonumber \\
&\overset{}{=}\int_0^{y} f_{\Vj}(v)  \!\!\! \PIjk \!\!\! \mathbb{P}\left[  R\ijk{\aN} (y) \leq \Vk    \right] dv, \nonumber
\end{align}
where (a) is from \eqref{eq:tau_rule}.
Here, for $y\to\infty$, the probability becomes equivalent to $\mathbb{P}\left[\MinP\;\big|\;a=\aN\right]$, which gives association probability in \eqref{eq:tau_Asso}.
In the strongest power association case, the main link has the strongest signal power, hence, 
the probability that main link is established as $\MinP$ and the main link distance is smaller than $y$ is given by
\begin{align}
&\mathbb{P}\left( \Yoa \leq y,\; \MinP \;|\;a=\aS\right) \label{eq:Sn_proof} \\
&\overset{(a)}{=} \int_0^{y} f_{\Vj}(v) \!\!\! \PIjk \!\!\! \mathbb{P} \left[ \frac{B_j}{v^{\alpha\Co}} \geq
\frac{B_k}{\left(\Vk\right)^{\alpha\To}}    \right] dv  \nonumber\\
&\overset{}{=}\int_0^{y} f_{\Vj}(v)  \!\!\! \PIjk \!\!\! \mathbb{P}\left[  R\ijk{\aS} (y) \leq \Vk    \right] dv,  \nonumber
\end{align}
where (a) is from \eqref{eq:tau_rule}.
Therefore, we derived the association probability \eqref{eq:tau_Asso} by $y\to\infty$. 

Finally, the \ac{CDF} of main link distance $\Yoa$ is given by
\begin{align}\label{eq:Cond_CDF}
&F_{\Yoa} (y) = 
\mathbb{P}\left(\Vj\leq y ,\; \MinP  \right)/ \AssoM,
\end{align}
which gives \eqref{eq:tau_PDF}.
\end{IEEEproof}


\section{Interference Analysis of MANs}

In this section, we analyze the Laplace transform of the interference in the \ac{MAN}.
In the \ac{MAN},
the interference to the $i$-layer receiver from the transmitters in the $k$-layer \ac{AN},
which have \ac{LoS} links ($\cb = \text{L}$) and \ac{NLoS} links ($\cb = \text{N}$) to the receiver is given by
%
%
\begin{align}
I\ikTo&=\sum\bxInt P_k G\To x^{-\alpha\To},\; \forall \Tin \label{eq:def_int}
\vspace{-3mm}
\end{align}
where $x$ is the link distance.
%
%
Let us define $\Emain$ as the event that using the rule $\tau$, 
 a $i$-layer receiver associates to a $j$-layer transmitter, and 
 their link distance is $y$ and channel environment is $c \in \{\text{L}, \text{N}\}$.
From the definition of the interference and the node association rules, the Laplace transform of the interference is given for the case of $\Emain$ in the following lemma.

	\begin{lemma}\label{L:LAPI}
	In the case of $\Emain$, the the Laplace transform of the interference from transmitters in the $k$-layer \ac{AN} under the channel environment $\cb$ is given by \eqref{eq:Lap_int}, which is presented on the top of {this} page,
	\begin{figure*}[!tbp]\vspace*{1pt}\begin{align}
		&\Lap{I\ikTo | \Emain}(s)=\exp\left[- 2\pi \lambda_{k,\text{Tx}} \left\{ \int_{\max \left(\chi\ijk{\tau}(y), \hik\right)}^\infty  x\rho\ikTo\left(x	\right)\left(1-\left(\frac{1}{1+\frac{sP_kx^{-\alpha\To}}{m\To}}\right)^{m\To} \right) dx  \right\}  \right]	\label{eq:Lap_int}
		\end{align}\hrulefill\normalsize\end{figure*}%
	where $\chi\ijk{\tau}(y)$ is given by
	\begin{align}\label{eq:chi_int}
	\chi\ijk{\tau}(y)=
	\begin{cases}
	\Rjk(y)& \text{for}\;o=\oR,   \\
	0 & \text{for}\;o=\oT,  \\
	\end{cases}
	\end{align}
	where $\Rjk(y)$ is defined in \eqref{eq:int_dis}.
\end{lemma}
	\begin{IEEEproof}
		%
%
In the case of $\Emain$, 
the Laplace transform of the interference is then given by
\begin{align}
&\Lap{I\ikTo | \Emain}(s) \label{eq:Lap_int_proof} \\
&\overset{}{=}\mathbb{E} \left[ \prod_{\bx\in\Phi\iktx\To}	\exp\left\{-sP_k G\To x\aCo		\right\}	 \;\biggr|\; \Emain \right] \nonumber \\
&\overset{(a)}{=}\mathbb{E}_{\Phi\iktx\To} \left[\!\prod_{\bx\in\Phi\iktx\To }\!	\left(	\!\frac{1}{1+\frac{sP_k x^{-\alpha\To}}{m\To}}\!\right)^{m\To} \biggr|\;\Emain\right].\nonumber
\end{align}
Here, (a) is obtained by averaging over the channel fading $G\To$, which gives the \ac{MGF} of Gamma distribution.
Since the density of interferer is $2 \pi x \lambda_{k,\text{Tx}} \rho\ikTo(x)$,
the \ac{PGFL} of \ac{PPP} for function $f(x)$ is obtained as\cite{HaeGan:09}
\begin{align}
&\mathbb{E}_{\Phi\iktx\To}   \left[ \prod_{\bx\in\Phi\iktx\To} f(\bx)\; \biggr|\; \Emain \right] \nonumber \\
&=\exp\left(-2\pi\lambda_{k,\text{Tx}} \int_{\chi}^\infty x(1-f(x))\rho\ikTo\left(x\right)dx\right) \label{eq:PGFL_PPP}
\end{align}
where $\chi=\max\left(\chi\ijk{\tau}, \hik\right)$ is the minimum distance bound for interferers.
When the node association rule $o=\oR$ is used, there is no interferer with shorter distance than \eqref{eq:int_dis} to the receiver since a receiver selects the nearest or the strongest transmitter, and we get $\ctau=\Rjk(y)$ in \eqref{eq:chi_int}. 
On the contrary, when $o=\oT$, a transmitter selects a receiver, so the locations of the interferers are independent with the location of the main link transmitter, and we get $\ctau=0$.
From \eqref{eq:Lap_int_proof} and \eqref{eq:PGFL_PPP}, we get the Laplace transform of the interference as \eqref{eq:Lap_int}.	

	\end{IEEEproof}

There is no closed form of \eqref{eq:Lap_int}. 
However, for the case of the interference from the transmitters in the same layer, i.e., $I_{ii}\Co$, we obtain the Laplace transform of the interference in a closed form as in the following corollary.

\begin{corollary}\label{cor:LAP_clo}
The Laplace transform of the interference from transmitters in the $i$-layer to the receiver in the $i$-layer is given by \eqref{eq:Lap_int_clo} when $sP_i\chi\ijk{\tau}(y)\aTo<1$ and $m^{(\text{L})}=m^{(\text{N})}=1$,
	\begin{figure*}[!tbp]\vspace*{1pt}\begin{align}\label{eq:Lap_int_clo}
	\Lap{I_{ii}\To | \varepsilon_{ii, \tau}\Co(y)}(s)=	
	&\begin{cases}
	\exp  \left[2\pi\lambda_{i,\text{Tx}} \sum_{n=1}^\infty \left(-s P_i \right)^n \eta^{n\alpha\To-2}\Gamma \left(2-n\alpha\To, \eta\chi_{j,i,\tau}^{(c,\cb)} (y)\right) \right] & (\cb)=(\text{L}) \\
	\exp  \left[2\pi\lambda_{i,\text{Tx}} \sum_{n=1}^\infty \left(-s P_i \right)^n \left\{\frac{\left(\chi_{j,k,\tau}^{(c,\cb)}\right)^{2-n\alpha^{(\cb)}}}{n\alpha^{(\cb)}-2}-\eta^{n\alpha\To-2}\Gamma \left(2-n\alpha\To, \eta\chi_{j,i,\tau}^{(c,\cb)} (y) \right) \right\}\right]  &(\cb)=(\text{N})  \\
	\end{cases}
	\end{align}\hrulefill\normalsize\end{figure*}%
where
\begin{align}
\eta&=-\sqrt{\mu\nu}\ln\left(1-\exp \left[ -\frac{h_i^2}{2\xi^2}\right]\right).
\end{align}
\end{corollary}
\begin{IEEEproof}
From Lemma~\ref{L:LAPI}, the Laplace transform of the interference in the \ac{LoS} environment, $I_{ii}\coL$, is given by
\begin{align}
&\Lap{I_{ii}\coL | \Emain}(s) \nonumber\\
\overset{}{=}&\exp\left\{- 2\pi \lambda_{i,\text{Tx}}  \int_\chi^\infty x e^{-\eta x}  \left(1-\left(\frac{1}{1+sP_i x\acoL}\right) \right) dx    \right\} \nonumber \\
\overset{(a)}{=}&\exp\left\{ 2\pi \lambda_{i,\text{Tx}}  \int_\chi^\infty x e^{-\eta x}  \sum_{n=1}^\infty \left( -s P_i x\acoL \right)^ndx    \right\} \label{eq:Lap_int_clo_L_proof1} \\\
\overset{}{=}&\exp\left\{ 2\pi \lambda_{i,\text{Tx}} \sum_{n=1}^\infty \left(\left( -s P_i \right)^n \int_\chi^\infty x^{1-n\alpha^{(\text{L})}} e^{-\eta x}   dx\right)     \right\}  \nonumber\\
\overset{(b)}{=}&\exp\left\{ 2\pi \lambda_{i,\text{Tx}} \right. \times\nonumber  \\
&\left. \sum_{n=1}^\infty \left(\left( -s P_i \right)^n \eta^{n\alpha^{(\text{L})}-2}\int_{\eta\chi}^\infty t^{1-n\alpha^{(\text{L})}} e^{-t}   dt  \right) \right\},   \label{eq:Lap_int_clo_L_proof2}
\end{align}
where $\chi={\max \left( \chi_{j,i,\tau}^{(c,\cb)} (y) , h_{ii}\right)}= \chi_{j,i,\tau}^{(c,\cb)}(y)$.
Here, 
(a) follows from the Taylor series $1/(1+x)=\sum_{n=0}^\infty (-x)^n$, which is convergent for $|x|<1$, so, \eqref{eq:Lap_int_clo_L_proof1} is convergent for $s P_i x\aCo<1$, and (b) follows from integration by substitution $\eta x=t$.
In \eqref{eq:Lap_int_clo_L_proof2}, by definition of the upper incomplete gamma function $\Gamma(x,y)=\int_y^\infty t^{x-1} e^{-t} dt$, we get the upper part of \eqref{eq:Lap_int_clo}.
In a similar way, the Laplace transform of the interference in the \ac{NLoS} environment, $I_{ii}\coN$, is given by
\begin{align}\label{eq:Lap_int_clo_L_proof}
&\Lap{I_{ii}\coN | \Emain}(s) \\
&\overset{}{=}\exp\left[ 2\pi \lambda_{i,\text{Tx}}  \int_\chi^\infty x (1-e^{-\eta x})  \sum_{n=1}^\infty \left( -s P_i x\acoN  \right)^ndx    \right] \nonumber\\
&=\exp  \left[2\pi\lambda_{i,\text{Tx}} \sum_{n=1}^\infty \left(-s P_i \right)^n \left(\frac{\left(\chi_{j,k,\tau}^{(c,\text{N})}\right)^{2-n\alpha\coN}}{n\alpha\coN-2}\right) \right]\times \nonumber \\
& \exp  \left[ -2\pi\lambda_{i,\text{Tx}} \int_\chi^\infty x e^{-\eta x}  \sum_{n=1}^\infty \left( -s P_i x\acoN  \right)^ndx	\right]\nonumber
\end{align}
From \eqref{eq:Lap_int_clo_L_proof}, we get the lower part of \eqref{eq:Lap_int_clo}.
	\end{IEEEproof}


For the $i$-layer receiver, the sum of total interference and noise is defined as
\begin{align}
\mathcal{I}_i=\sum_{k\in\mathcal{K}, \Tin} I\iktx\To+\sigma^2 \label{eq:total_int}
\end{align}
where $\sigma^2$ is the noise power.
From the property of the Laplace transform, the Laplace transform of $\mathcal{I}_i$ is given by
\begin{align}
\Lap{\mathcal{I}_i \;|\; \Emain }(s)&\!=\!	\exp(-s\sigma^2)\!\prod_{k\in\setTx,\Tin}\!\Lap{I\ikTo \;|\; \Emain}(s).	\label{eq:Lap_total_int}
\end{align}


	\section{Performance Analysis of MANs}
In this section, we analyze the \ac{STP} and the \ac{ASE} of \ac{MAN} based on the Laplace transform of the interference.
In addition, we derive the upper bound of the optimal density that maximizes the \ac{STP} and the \ac{ASE}.

\subsection{STP and ASE Analysis}
In the event of $\Emain$, the \ac{STP} is defined using \ac{SINR} as
\begin{align}
\poa{\tau}(y)&=\mathbb{P}\left[\text{SINR}\ijCo(y)	>\beta_{ij} \;|\; \Emain \right],			\label{eq:def_STP_c}
\end{align}
where 
\begin{align}
\text{SINR}\ijCo(y)=P_j G\Co y\aCo/\mathcal{I}_i. \label{eq:def_SINR}
\end{align}
Here, $\beta_{ij}$ is the target \ac{SINR}, which is related to the target transmission rate between a $i$-layer receiver and a $j$-layer transmitter.
In addition, the definition of \ac{ASE} is given as the sum of the maximum average data rates per unit bandwidth per unit area for a specified bit error rate \cite{AloGol:99,VikChaJef:09}. 
We assume the number of the communication links in the unit area depends on the number of the transmitters and the number of the receivers for $o=\oT$ and $o=\oR$, respectively.
Therefore, when $o=\oR$, we define the \ac{ASE} as the data rate multiplied with the density of the receiver. On the contrary, when $o=\oT$, we define the \ac{ASE} as the data rate multiplied with the density of the receiver.
Here, the data rate is $\log(1+\beta_{ij})$ when the communication succeeds, and $0$ when the communication is failed.
Therefore, the \ac{STP} and the \ac{ASE} of the $k$-layer in the \ac{MAN} is derived as the following Lemma.

	%
	\begin{lemma}\label{L:r_Performance}
		Using the node association rule $\tau$, the \ac{STP} and the \ac{ASE} of the $k$-layer in the \ac{MAN} is given as
\begin{align}
&\STP{k} \label{eq:STPk}=
\begin{cases}
\sum_{	j \in \mathcal{K},\Cin} \frS&\text{for}\; o=\oR,\\
\sum_{	i \in \mathcal{K},\Cin} \ftS& \text{for}\; o=\oT,
\end{cases}\\
&\ASE{k} = \label{eq:ASE}
\begin{cases}
\sum_{	j \in \mathcal{K},\Cin} (\lambda_j-\!\lambda_{j,o}) R_{ij} \frS\; \text{for}\; o=\oR,\\
\sum_{	i \in \mathcal{K},\Cin} (\lambda_i-\!\lambda_{i,o}) R_{ij} \ftS\;\; \text{for}\; o=\oT.
\end{cases}
\end{align}
where $R_{ij}=\text{log}_2(1+\beta_{ij})$ and $\fS$ is given as
\begin{align}
\fS=\AssoM\int_{\hij}^\infty \poa{\tau}(y) f_{\Yoa}(y) dy.
\end{align}
Here, $\poa{\tau}(y)$ is presented as 
		\begin{align}
		&\poa{\tau}(y)
		=\left. \sum_{n=0}^{m\Co -1}\frac{(-s)^n}{n!}\frac{d^n}{d s^n}\Lap{\mathcal{I}_i| \Emain}  \left( s	\right) \right|_{s=\LapS},  \label{eq:comp_STP}\\
		&\LapS=\frac{m\Co\beta_{ij}}{P_j y\aCo}, \label{eq:Sof_STP}
		\end{align}
		%
		%
		where $\Lap{\mathcal{I}_i| \Emain}  \left( s	\right)$ is in \eqref{eq:Lap_total_int}.
		%
	\end{lemma}
	\begin{IEEEproof}
		From \eqref{eq:def_STP_c} and \eqref{eq:def_SINR}, 
		the \ac{STP} is given by
		%
		\begin{align}
		&\poa{\tau}(y)=\mathbb{P}\left[ G\Co 	> \frac{\beta_{ij} \mathcal{I}_i}{P_jy^{-\alpha\Co}}	\;\biggr|\;\Emain\right]\label{eq:comp_STP_proofx} \\
		&\overset{(a)}{=}\mathbb{E}_{\mathcal{I}_i}\left. \left[\sum_{n=0}^{m\Co -1} \frac{\left(s \mathcal{I}_i \right)^n}{n!} \exp\left(- s \mathcal{I}_i	\right) \;\biggr|\;	\Emain	\right]	\right|_{s=\LapS}		\nonumber
		\end{align}
		where (a) follows from the Gamma distribution of channel gain and the property of lower incomplete Gamma function.
		Notice that we derived \eqref{eq:Sof_STP} from (a).
		Using following property of the Laplace transform, 
		we obtain \eqref{eq:comp_STP}.
		\begin{align}
		\mathcal{L}_{\mathcal{I}_i}(s)=\mathbb{E}_{\mathcal{I}_i}\left[\exp\left(-s \mathcal{I}_i\right)\right],\nonumber \\
		(-\mathcal{I}_i)^n \Lap{\mathcal{I}_i}(s)=\frac{d^n}{d s^n}\Lap{\mathcal{I}_i}(s).
		\end{align}
	
		Therefore, from the \ac{PDF} of the main link distance and the association probability in Lemma~\ref{L:PDFM}, we obtain the \ac{STP} of node in the $k$-layer as \eqref{eq:STPk}.
		Furthermore, by the definition of the \ac{ASE}, which is the data rate multiplied with the density of the node in the $k$-layer, we obtain \ac{ASE} of $k$-layer \ac{AN} in the \ac{MAN} as \eqref{eq:ASE}.		
	\end{IEEEproof}


In the \ac{MAN}, the \ac{STP} and the \ac{ASE} of the network is given by
\begin{align}\label{eq:total_Perf}
\STP{\text{MAN}}&=
\begin{cases}
\sum_{i\in\mathcal{K}}\frac{\lambda_{i,\Rx}}{\lambda_{T,\Rx}}\STP{i} & \text{for}\; o=\oR,\\
\sum_{i\in\mathcal{K}}\frac{\lambda_{i,\Tx}}{\lambda_{T,\Tx}}\STP{i} & \text{for}\; o=\oT,
\end{cases}\\
\ASE{\text{MAN}}&=
\sum_{i\in\mathcal{K}}\ASE{i}.
\end{align} 
Note that the \ac{STP} of the \ac{MAN} is defined as the average \ac{STP} of the receivers(transmitters), whereas the \ac{ASE} of the \ac{MAN} is defined as the total amount of data rate per unit frequency per area.

The \ac{STP} and the \ac{ASE} have multiple integral which makes evaluation hard, however, the integral in the Laplace transform can be removed under the condition, which explained in the following remark.
\begin{remark}
	In Lemma~\ref{L:r_Performance}, the Laplace transform of the interference from the same layer of AN can be replaced with \eqref{eq:Lap_int_clo} when $\tau=\text{rs}$, $\beta_{ij}<1$, $m^{(\text{L})}=m^{(\text{N})}=1$, and $B_k=P_k$, since the conditions in Corollary~\ref{cor:LAP_clo} are satisfied. 
\end{remark}
\begin{IEEEproof}
In the event of $\Emain$, using the node association $\tau=\text{rs}$, $\ctau=\Rjk$. 
Furthermore, when $B_k=P_k$,  $\frac{P_k x\aTo}{P_j y\aCo}<1$ for all $\Rjk \leq x$.
Therefore, $\LapS P_k x\To =\beta_{ij}\frac{P_k x\aTo}{P_j y\aCo}<1$, which satisfy the condition in Corollary~\ref{cor:LAP_clo}.
\end{IEEEproof}

\subsection{Upper Bound of Optimal Density}

In the design of the \ac{MAN}, it is important to optimize the densities of transmitters in order to maximize the \ac{STP} and the \ac{ASE}.\footnote{Note that the optimal density of the receiver is trivial to derive.
From our analysis, $\mathcal{P}_{k,\oR a}$ is independent with the density of the receiver. Furthermore, $\mathcal{P}_{k,\oT a}$ and $\ASE{k}$ increase with the density of the receiver.}
However, it is hard to present the \ac{STP} or the \ac{ASE} in a closed form, so, consequently, hard to obtain the optimal densities. 
Nevertheless, we derived the upper bound of the optimal density that maximizes the \ac{STP} and the \ac{ASE} in the following corollary.

	\begin{corollary}\label{cor:upper_bound}
	When $j$-layer transmitters communicate to $k$-layer receivers
	and  the channel coefficient is $m\coN=m\coL=1$, 
		the upper bound of the optimal transmitter density in the $j$-layer that maximizes the \ac{STP}, $\Lb{\STP{k}}$, 
		and that maximizes the \ac{ASE}, $\Lb{\ASE{k}}$, of the $k$-layer are respectively given by  
		\begin{align}
		\Lb{\STP{k}}&= \label{eq:STP_b} 
		\begin{cases}
		\frac{1}{2\pi  \epsilon_{kj} } & \text{for}\; o=\oR,\\
		0 & \text{for}\; o=\oT,
		\end{cases}\\
		\Lb{\ASE{k}}&= \label{eq:ASE_b}
		\begin{cases}
		\frac{1}{2\pi  \epsilon_{kj} } & \text{for}\; o=\oR,\\
		0 & \text{for}\; o=\oT\;\text{and}\;k\neq j, \\
		\max_{i\in\mathcal{K}}\left[\frac{1}{2\pi  \epsilon_{ki} }\right] & \text{for}\; o=\oT\;\text{and}\;k= j .
		\end{cases}
		\end{align}
where $\epsilon_{ij}$ is 
		\begin{align}
		&\epsilon_{ij}  =\label{eq:epsil} \\
		& \int_{h_{ij}}^\infty x\left(1-\frac{\rho_{ij}\coL (x)}{1+\beta_{ij} h_{ij}^{\alpha^{(\text{L})}} x\acoL}-\frac{\rho_{ij}\coN(x)}{1+\beta_{ij}  h_{ij}^{\alpha^{(\text{L})}} x\acoN}  \right)dx.  \nonumber
		\end{align}
	\end{corollary}
	\begin{IEEEproof}
		See Appendix~\ref{app:upper_bound}.
\end{IEEEproof}

Note that the optimal density exists due to the trade-off: 
increasing the transmitter density increases the ASE or STP due to the shorter link distance and the increasing number of communication links in the network, while also decreases the ASE or STP due to the larger interference.
However, when $o=\oT$, the \ac{STP} always decreases with the transmitter density since the main link distance distribution is determined by the receiver density not by the transmitter density, which gives the optimal transmitter density as zero.
Furthermore, {when $o=\oT$ and $j\neq k$, the \ac{ASE} of $j$-layer always decreases with the $k$-layer transmitter density since neither the main link distance nor the number of communication links in the $j$-layer depend on the $k$-layer transmitter density.} 
As the optimal density of the transmitters is hard to be presented, we need to use a certain search algorithm such as the exhaustive search.
In that case, this upper bound can be usefully used to determine the search range.
In addition, as shown in Corollary~\ref{cor:upper_bound}, 
the optimal transmitter density bound can be determined for each layer independently as it is not affected by other layer transmitter densities.
From Corollary~\ref{cor:upper_bound}, we can also see the following tendency of the upper bound.
\begin{corollary}\label{cor:dec}
The upper bound of the optimal transmitter densities $\Lb{\STP{k}}$ and $\Lb{\ASE{k}}$, are non-increasing function of $\hij$ under the conditions of $\beta_{ij} \hij^{\alpha\coL}>1$ and $h_{ij}>1$, since $\epsilon_{ij}$ increases with $\hij$.
\end{corollary}
\begin{IEEEproof}
	See Appendix~\ref{app:dec}.
\end{IEEEproof}
Note that the condition $\beta_{ij} \hij^{\alpha\coL}>1$ and $\hij>1$ are conditions, which are generally satisfied in \ac{UAV} communications.
From Corollary~\ref{cor:dec}, we can see that 
as the altitude difference between the transmitter and the receiver $h_{ij}$ increases, the optimal transmitter density bound becomes smaller, which will be almost zero for large altitude difference.

\begin{table}
	\caption{Simulation Parameters} \label{table:parameters}
	\begin{center}
		\rowcolors{2}
		{cyan!15!}{}
		\renewcommand{\arraystretch}{1.3}
		\begin{tabular}{c c c c} 
			\hline
			{\bf Parameter} & {\bf Value} & {\bf Parameter} & {\bf Value}
			\\
			\midrule
			\hline
			$(m\coL, m\coN)$ & (1, 1) & $(\mu,\nu,\xi)$ & (0.5, $3\times 10^{-4}$, 20) \\ \addlinespace
			$(\alpha\coL, \alpha\coN)$ & (2.5, 3.5) &  $(a,b)$ & (12.0910, 0.1139)\\ \addlinespace
			$a$ & $a=\aS$ & $\beta_k$ & 0.7\\ \addlinespace
			$(B_k, P_k)$  &  1 &  Receiver layer & $0$\\ \addlinespace
			$\sigma^2$ & 0  &
			Transmitter layer & $j, k$\\ \addlinespace
			\hline
		\end{tabular}
	\end{center}\vspace{-0.63cm}
\end{table}%

\section{Numerical Results}\label{sec:Num}
In this section, we present the \ac{STP} and the \ac{ASE} of the \ac{MAN} for
the receiver-oriented and the transmitter-oriented association cases. 
For the numerical results, we consider the interference-limited environment, i.e., $\sigma^2=0$, 
in order to clarify the results. 
We use the ground layer, i.e., $0$-layer, and $i$-layer as receivers' layer, $j$-layer and $k$-layer as the transmitters' layers, and omit the subscripts $\Rx$ and $\Tx$ for the simplicity, e.g., $\lambda_0=\lambda_{0, \Rx}$ and $\lambda_j=\lambda_{j, \Tx}$. 
Furthermore, we omit the subscript in the total transmitter density, i.e., $\lambda_j+\lambda_k=\lambda_T$ instead of $\lambda_{T, \Tx}$.
Simulation parameters for our numerical results summarized in Table.~\ref{table:parameters}, where Fig.~\ref{fig:R_O_HM} uses $m\coL=3$ and the nearest distance association, i.e., $a=\aN$, and Fig.~\ref{fig:R_O_Diff_LoSP} uses the $i$-layer as the receivers' layer.

\subsection{Receiver-oriented Association Case}\label{sub:Num_Rec}

\begin{figure}[t!]
	\begin{center}
		{
			\psfrag{AAAAAAAAAAA}[][][0.9]{$m\coL=3$}
			\psfrag{AAAAAAAAAAB}[][][0.9]{$m\coL=1$}
			\psfrag{AAAAAAAA}[][0.9]{$\tau=\text{rs}$}
			\psfrag{AAAAAAAB}[][0.9]{$\tau=\text{rn}$}
			\psfrag{AAAAAAAC}[][0.9]{$\rho_{0j}\coL(x)=1$}
			\psfrag{EAAAAAAAAAA}[][][0.8]{$h\jtxn$ (m)}
			\psfrag{FAAAAAAAAAA}[][][0.8]{Successful Transmission Probability (STP)}
			\includegraphics[width=1.00\columnwidth]{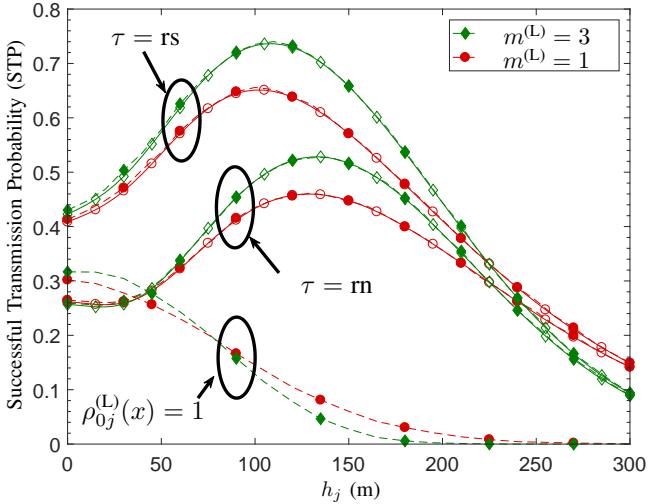}
			\vspace{-8mm}
		}
	\end{center}
\caption{
	\ac{STP} of the single layer \ac{AN} according to the transmitter altitude $h\jtxn$ with different \ac{LoS} coefficientes $m\coL$ when $\lambda\jtxn=10^{-5}$, $o=\oR$.
}   \label{fig:R_O_HM}
\end{figure}
\begin{figure}[t!]
	\begin{center}
		{
			\psfrag{AAAAAAAAAAAAAA}[][][0.9]{$h_i=0$}
			\psfrag{AAAAAAAAAAAAAB}[][][0.9]{$h_i=10$}
			\psfrag{AAAAAAAAAAAAAC}[][][0.9]{$h_i=20$}
			\psfrag{AAAAAAAAAAAAAD}[][][0.9]{$h_i=30$}
			\psfrag{AAAAAAAAAAAAAE}[][][0.9]{$h_j=h_i$}
			\psfrag{EAAAAAAAAAA}[][][0.8]{$h\jtxn-h_i$ (m)}
			\psfrag{FAAAAAAAAAA}[][][0.8]{Successful Transmission Probability (STP)}
			\includegraphics[width=1.00\columnwidth]{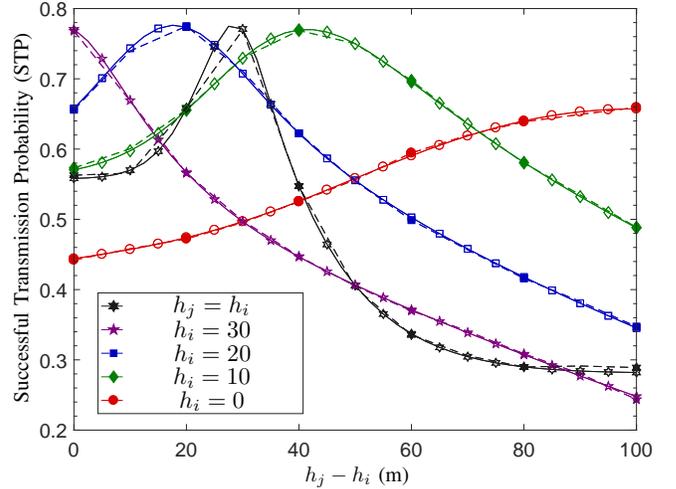}
			\vspace{-8mm}
		}
	\end{center}
\caption{
	\ac{STP} of the single layer \ac{AN} according to the difference of altitude between layers $h_j-h_i$ with different receiver altitudes $h_i$ when $\lambda_{j}=10^{-5}$ and $\tau=\oR\aS$.
}   \label{fig:R_O_Diff_LoSP}
\end{figure}
\begin{figure}[t!]
	\begin{center}
		{
			\psfrag{EAAAAAAAAAA}[][][0.8]{$h\jtxn$ (m)}
			\psfrag{FAAAAAAAAAA}[][][0.8]{$\lambda\jtxn$ (number/$\text{m}^2$)}
			\includegraphics[width=1.00\columnwidth]{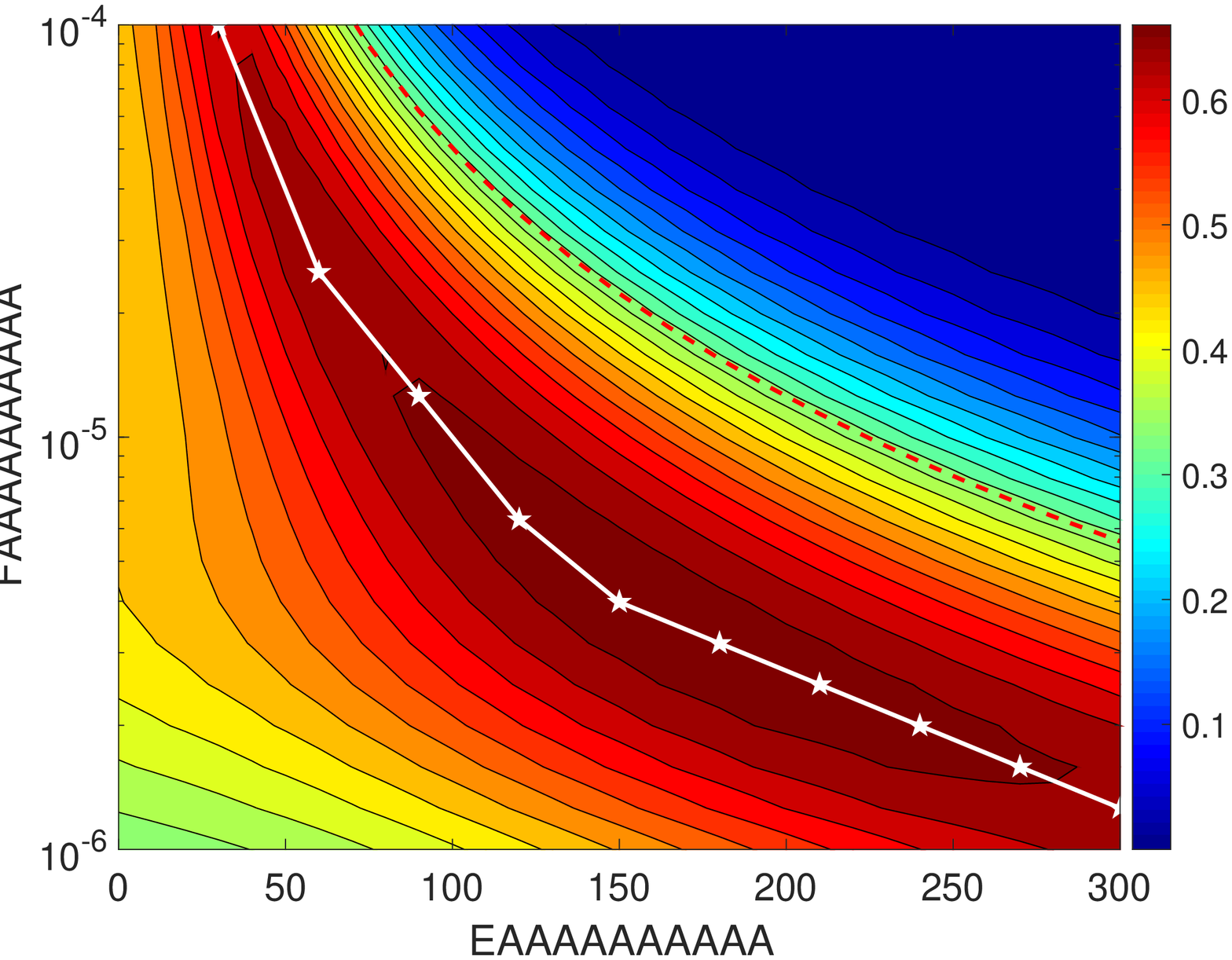}
			\vspace{-8mm}
		}
	\end{center}
\caption{
	\ac{STP} of the single layer \ac{AN} as functions of the transmitter density $\lambda\jtxn$ and the transmitter altitude $h\jtxn$ when $\tau=\oR\aS$. 
	A solid line with stars presents the optimal density and a dotted line presents the upper bound of the optimal density.
}   \label{fig:R_O_STP}
\end{figure}
\begin{figure}[t!]
	\begin{center}
		{
			\psfrag{EAAAAAAAAAA}[][][0.8]{$\lambda_{j}$ (number/$\text{m}^2$)}
			\psfrag{FAAAAAAAAAA}[][][0.8]{$\lambda_{k}$ (number/$\text{m}^2$)}
			\psfrag{AAAAAA}[][][0.7]{$\lambda_{T}=10^{-6}\;$}
			\psfrag{AAAAAB}[][][0.7]{$\lambda_{T}=10^{-5.3}$}
			\psfrag{AAAAAC}[][][0.7]{$\lambda_{T}=10^{-4.6}$}
			\includegraphics[width=1.00\columnwidth]{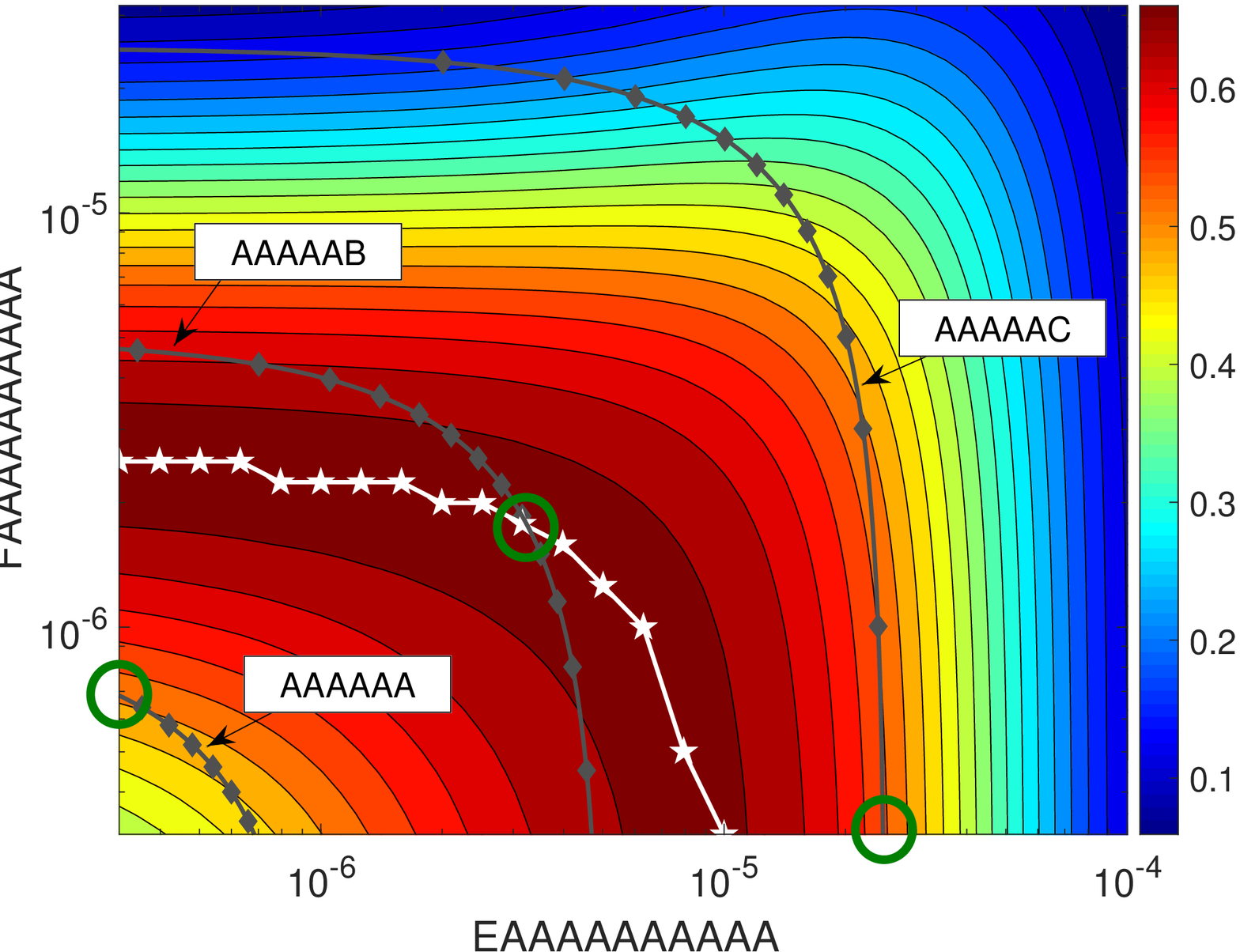}
			\vspace{-8mm}
		}
	\end{center}
\caption{
	\ac{STP} of the two layer \ac{MAN} as functions of the $j$-layer transmitter density $\lambda\jtxn$ and the $k$-layer transmitter density $\lambda\ktxn$ when $h\jtxn=100$, $h\ktxn=200$, and $\tau=\oR\aS$. 
	A line with stars presents the optimal $j$-layer transmitter density and a line with diamonds presents the area that have the same total density $\lambda_T$.
}   \label{fig:R_O_2Layer_STP}
\end{figure}

In this subsection, we show the \ac{STP} of the \ac{MAN} when the receiver-oriented association is considered. 
We omit the \ac{ASE} results since the \ac{ASE} is a multiplication of the \ac{STP} with the receiver density when $o=\oR$, which gives the same tendency with the \ac{STP}.
To show the effects of network parameters on the performance more clearly, we first show the performance for a single layer \ac{AN} case in Figs.~\ref{fig:R_O_HM}-\ref{fig:R_O_STP}, and then provide the performance for a two layer \ac{MAN} case in Fig.~\ref{fig:R_O_2Layer_STP}. 

Figure~\ref{fig:R_O_HM} shows the \ac{STP} of the single layer \ac{AN} (i.e., the $j$-layer) as a function of the altitude of the layer $h\jtxn$ for different values of channel coefficient $m\coL=\{1,3\}$ and two node association rules, i.e., the strongest power ($\tau=\text{rs}$) and the nearest distance ($\tau=\text{rn}$) associations. Here, the density of the transmitter is $\lambda\jtxn=10^{-5} $ [nodes/$\text{m}^2$].
The \ac{LoS} probability in \eqref{eq:LoS_A2G} is used for this figure, and we also provide the results with $\rho_{0j}\coL=1$ (i.e., the case that always assumes \ac{LoS} link) to show the effect of the \ac{LoS} probability consideration.
Simulation results are obtained from Monte Carlo simulation which are presented by the dashed lines with filled markers, 
while analysis results are presented by the solid lines with unfilled markers, which fit well with the simulation results.

From Fig.~\ref{fig:R_O_HM}, we observe the existence of the optimal altitude of the transmitter layer due to the trade-off by the altitude on the \ac{STP}: 
as the altitude of the transmitter increases, 
the \ac{LoS} probability of the main link also increases, which results in higher \ac{STP},
while both the \ac{LoS} probability of the interference link and the main link distance increase, 
which lowers \ac{STP}.
However, when the \ac{LoS} probability is $\rho_{0j}\coL=1$ and not changed with the altitude, the \ac{STP} only decreases since the main link distance increases with the altitude.

In addition, we observe the effect of the \ac{LoS} coefficient $m\coL$ on the \ac{STP}, 
which gives higher \ac{STP} at low altitude region (e.g., $h_j <200$) 
and gives lower \ac{STP} at high altitude region (e.g., $h_j >230$). 
For the Nakagami-m fading, the larger coefficient $m\Co$ gives less chance to have the smaller channel gain.
At the low altitude region, the main link is mostly \ac{LoS} while the interference is \ac{NLoS} that gives the higher \ac{SINR} with the larger \ac{LoS} coefficient $m\coL$. 
On the contrary, at the high altitude region, the interference has more \ac{LoS} links that gives the lower \ac{SINR} with the larger \ac{LoS} coefficient $m\coL$.

Figure~\ref{fig:R_O_Diff_LoSP} depicts the \ac{STP} of the single-layer \ac{AN}\footnote{Even though we use two \acp{AN}, $i$ and $j$-layer, we regard it as the single-layer \ac{AN} since only one layer acts as the transmitter and the receiver.} as a function of the altitude difference between the $j$-layer (i.e., transmitter layer) and the $i$-layer (i.e., receiver layer), $h\jtxn-h_i$, for different altitudes of the $i$-layer, $h_i=\{0, 10, 20, 30\}$, when $h\jtxn>h_i$.
Here, the strongest power association ($\tau=\text{rs}$) is used and $\lambda\jtxn=10^{-5}$. 
The simulation results are presented by the dashed lines with filled markers, while analysis results are presented by the solid lines with unfilled markers. 
For the analysis results, we use Lemma 2 
for $h_i=\{0, 10, 20, 30\}$ and Corollary 1  
for the case of $h_i=h\jtxn$, and show that analysis results match well with the simulation results.\footnote{Although the closed form Laplace transform contains the infinite summation in Lemma 2 and we use partial summation, i.e., $\sum_{n=1}^{10}$, instead of $\sum_{n=1}^{\infty}$, we show high coincidence with simulation results since the partial summation converges to the infinite sum with a bearable error.}

From Fig.~\ref{fig:R_O_Diff_LoSP}, we observe that the optimal altitude difference $(h\jtxn-h_i)^*$ that maximizes the \ac{STP} exists and decreases with the h $h_i$.
When $h_i$ is large, the \ac{LoS} probability of the main link and the interfering links are high, hence, the smaller distance gives the higher \ac{STP} that reduce optimal $h\jtxn-h_i$.
Therefore, the difference between altitues should be smaller when the communication between different \acp{AN} in high altitude is considered. 
Considering $h_i=h\jtxn$, which is the same with the communication between nodes in the same layer, optimal altitude of layer $h_i$ exist since the \ac{LoS} probability of the main link and the interfering links increases with $h_i$. 
At low altitude region (e.g., $h_i=0$), the channel is mostly \ac{NLoS}, at high altitude region (e.g., $h_i=100$), contrary, the channel is mostly \ac{LoS}.
In between low and high (e.g., $h_i=30$), the \ac{LoS} probability is high when the smaller horizontal distance is considered, therefore, the main link is under \ac{LoS} channel whereas the interfering links are under \ac{NLoS} channel, that gives the higher \ac{STP} compared with the low and high altitude.

Figure~\ref{fig:R_O_STP} shows the \ac{STP} of the single-layer \ac{AN} as functions of the transmitter density $\lambda\jtxn$ and the transmitter altitude $h\jtxn$ when $\tau=\oR\aS$.
	We present the optimal density $\lambda_j^{*}$ that maximizes the \ac{STP} using a solid line with stars and the upper bound of the optimal density, obtained from Corollary~\ref{cor:upper_bound}, as a dashed line. 
	In addition, we observe the existence of the optimal density since 
	as the density of the transmitter increases, 
	the main link distance decreases and the \ac{LoS} probability of the main link increases, which results in higher \ac{STP},
	while both the interfering nodes and the \ac{LoS} probability of the interferers increase, which results in lower \ac{STP}.

Furthermore, by comparing the optimal density and the upper bound of the optimal density,
	we notice that their trends according to the altitude $h_j$ are similar. Specifically, both the optimal density and its upper bound decrease with 
	the altitude $h_j$ as proven in Corollary~\ref{cor:dec}.
	Although the difference between the optimal density and its upper bound is not small, 
	the upper bound can play an important
to find the optimal density by restricting the searching range, e.g., exhaustive searching starting from the upper bound.

Figure~\ref{fig:R_O_2Layer_STP} shows the \ac{STP} of the two layer \ac{MAN} as functions of the density of  $j$-layer transmitters $\lambda\jtxn$ and the density of $k$-layer transmitters $\lambda\ktxn$, 
	when $h\jtxn=100$, $h\ktxn=200$, and $\tau=\oR\aS$.
	The line marked with stars shows the optimal transmitter density of the $k$-layer $\lambda\ktxn^{*}$ for different values of $\lambda\jtxn$. We can see that $\lambda\ktxn^{*}$ decreases as $\lambda\jtxn$ increases. This is because the larger interference from the $j$-layer that makes the density of other interfering layer to decrease (i.e., $k$-lyaer), for the optimal density.
	The lines marked with diamonds show the cases of having the given values of the total density, i.e., 
$\lambda\jtxn +\lambda\ktxn = \lambda_T$,
and the points of circles shows the optimal densities $(\lambda\jtxn,\lambda\ktxn)^{*}$ for each cases of $\lambda_T$. We can see that when $\lambda_T$ is large (e.g., $\lambda_T = 10^{-4.6}$), having all transmitters in the layer with lower altitude (i.e., the $j$-layer) can achieve higher \ac{STP},  
while for small $\lambda_T$ (e.g., $\lambda_T = 10^{-6}$), having all transmitters in the layer with higher altitude (i.e., the $k$-layer) achieves higher \ac{STP}. However, when $\lambda_{T}$ is neither large or small, e.g., $\lambda_{T}=10^{-5.3}$, having transmitters in multiple layers, i.e., both $j$ and $k$-layers can be better in terms of the \ac{STP}.

%

\subsection{Transmitter-oriented Association Case}

\begin{figure}[t!]
		\begin{center}
			{
				\psfrag{EAAAAAAAAAA}[][][0.8]{$h\jtxn$ (m)}
				\psfrag{FAAAAAAAAAA}[][][0.8]{$\lambda\jtxn$ (number/$\text{m}^2$)}
				\includegraphics[width=1.00\columnwidth]{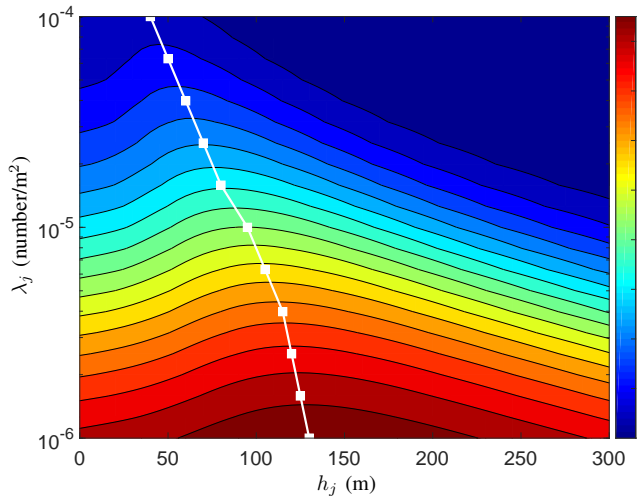}
				\vspace{-8mm}
			}
		\end{center}
\caption{
	\ac{STP} of the single layer \ac{AN} as functions of the transmitter density $\lambda\jtxn$ and the transmitter altitude $h\jtxn$ when $\tau=\oT\aS$.
	A solid line with squars presents the optimal altitude.
}   \label{fig:T_O_STP}
	\end{figure}

\begin{figure}[t!]
		\begin{center}
			{
				\psfrag{EAAAAAAAAAA}[][][0.8]{$h\jtxn$ (m)}
				\psfrag{FAAAAAAAAAA}[][][0.8]{$\lambda\jtxn$ (number/$\text{m}^2$)}
				\includegraphics[width=1.00\columnwidth]{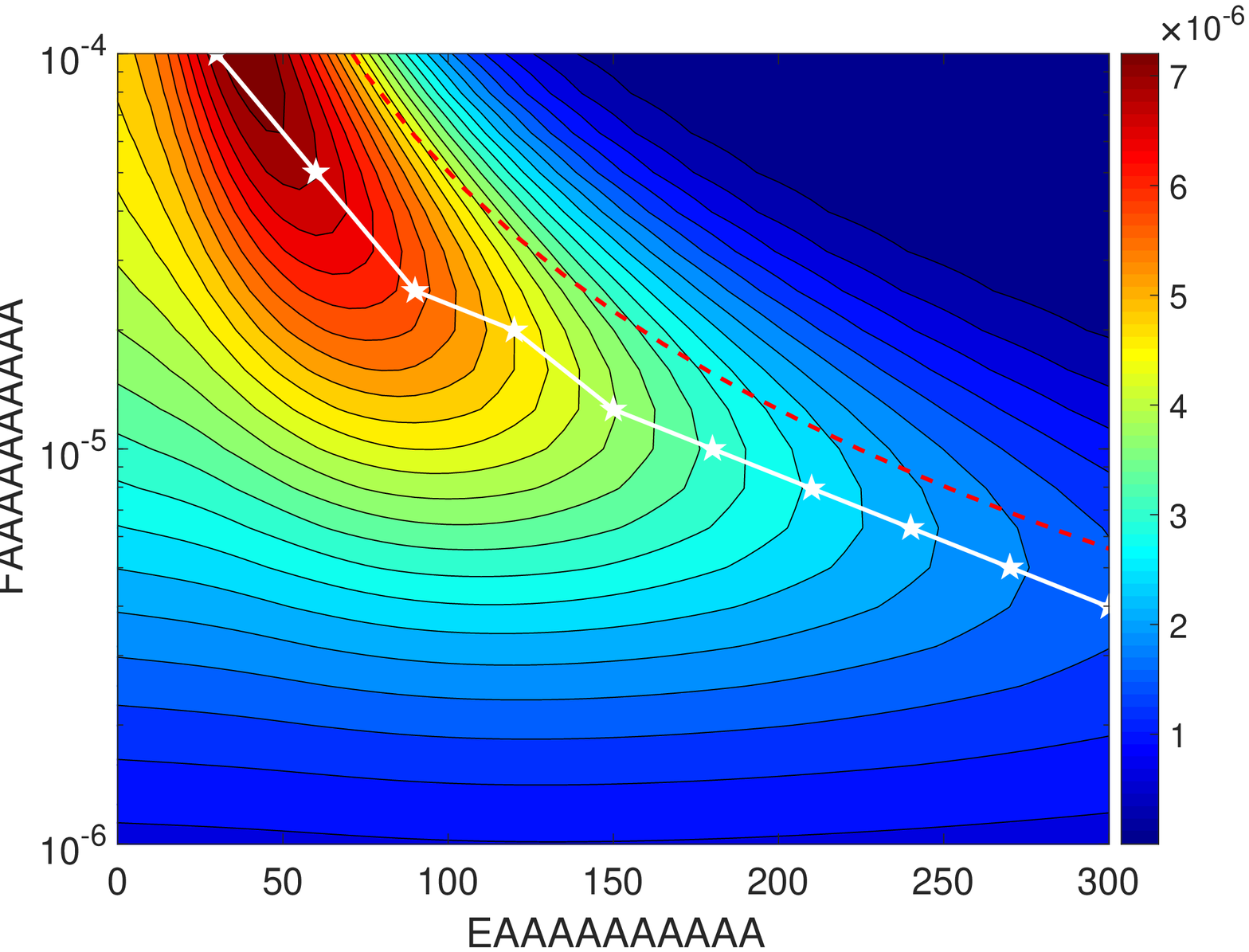}
				\vspace{-8mm}
			}
		\end{center}
\caption{
	\ac{ASE} of the single layer \ac{AN} as functions of the transmitter density $\lambda\jtxn$ and the transmitter altitude $h\jtxn$ when $\tau=\oT\aS$.
	A solid line with stars presents the optimal density and a dotted line presents the upper bound of the optimal density.
}   \label{fig:T_O_ASE}
	\end{figure}

\begin{figure}[t!]
\begin{center}
	{
				\psfrag{EAAAAAAAAAA}[][][0.8]{$\lambda\jtxn$ (number/$\text{m}^2$)}
\psfrag{FAAAAAAAAAA}[][][0.8]{$\lambda\ktxn$ (number/$\text{m}^2$)}
\psfrag{AAAAAA}[][][0.7]{$\lambda_{T}=10^{-6}\;$}
\psfrag{AAAAAB}[][][0.7]{$\lambda_{T}=10^{-5}\;$}
\psfrag{AAAAAC}[][][0.7]{$\lambda_{T}=10^{-4.5}$}
		\includegraphics[width=1.00\columnwidth]{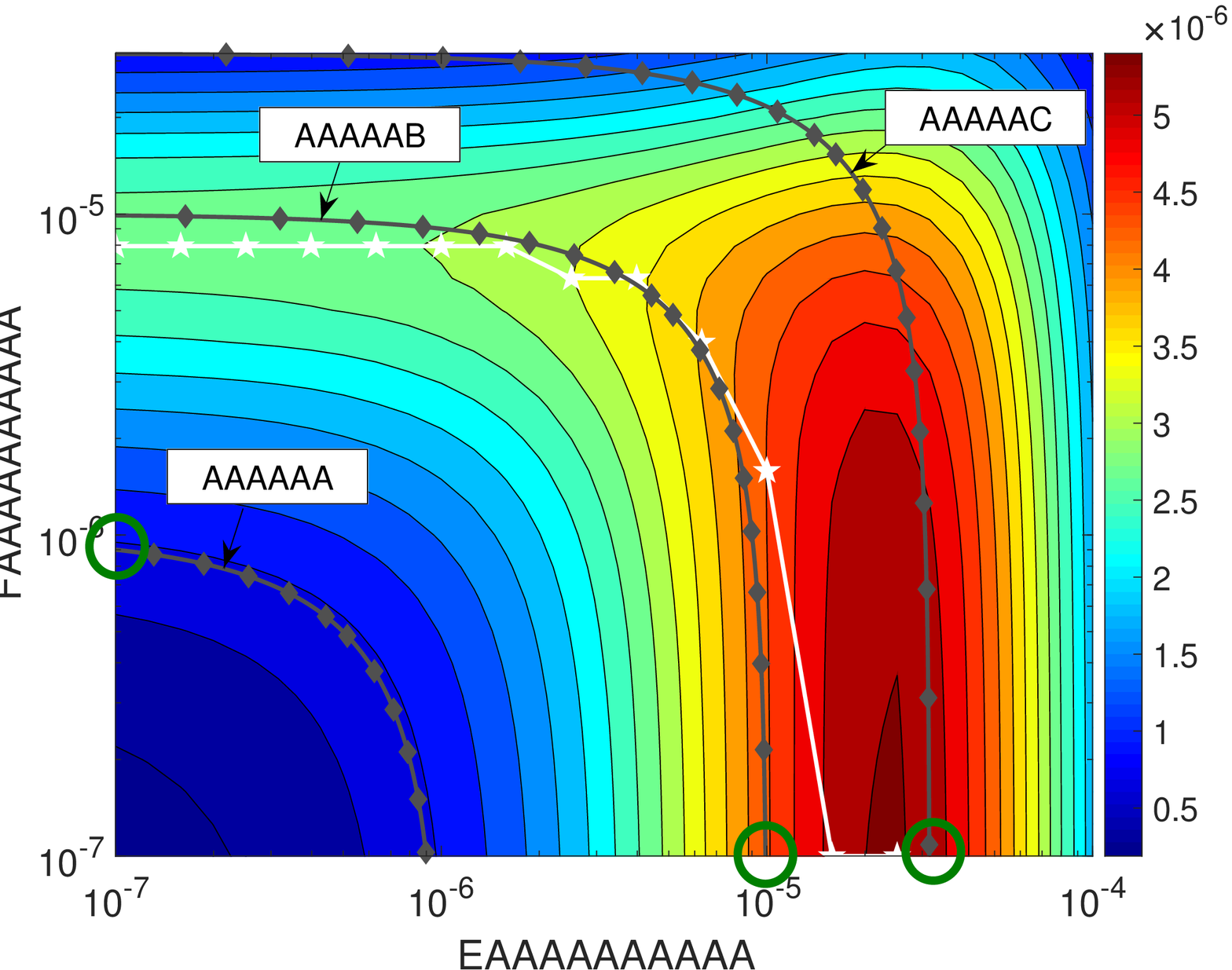}
		\vspace{-8mm}
	}
\end{center}
\caption{
	\ac{ASE} of the two layer \ac{MAN} as functions of the $j$-layer transmitter density $\lambda\jtxn$ and the $k$-layer transmitter density $\lambda\ktxn$ when 
	$h\jtxn=100$, $h\ktxn=200$, and $\tau=\oT\aS$.
	A line with stars presents the optimal $j$-layer transmitter density and a line with diamonds presents the area that have the same total density $\lambda_T$.
}   \label{fig:T_O_2Layer_ASE}
\end{figure}

\begin{figure}[t!]
	\begin{center}
		{
			\psfrag{AAAAAAAAAAAAA}[l][l][0.8]{$\lambda_{T}=10^{-6}\;\;$}
			\psfrag{AAAAAAAAAAAAB}[l][l][0.8]{$\lambda_{T}=1.4\times10^{-6}$}
			\psfrag{AAAAAAAAAAAAC}[l][l][0.8]{$\lambda_{T}=3.2\times10^{-6}$}
			\psfrag{AAAAAAAAAAAAD}[l][l][0.8]{$\lambda_{T}=10^{-5}\;\;$}
			\psfrag{AAAAAAAAAAAA}[][][0.8]{Ratio of $j$-layer transmitters in \ac{MAN}, $\lambda\jtxn/\lambda_{T}$}
			\psfrag{AAAAAAAAAAAB}[][][0.8]{Normalized ASE}
			\includegraphics[width=1.00\columnwidth]{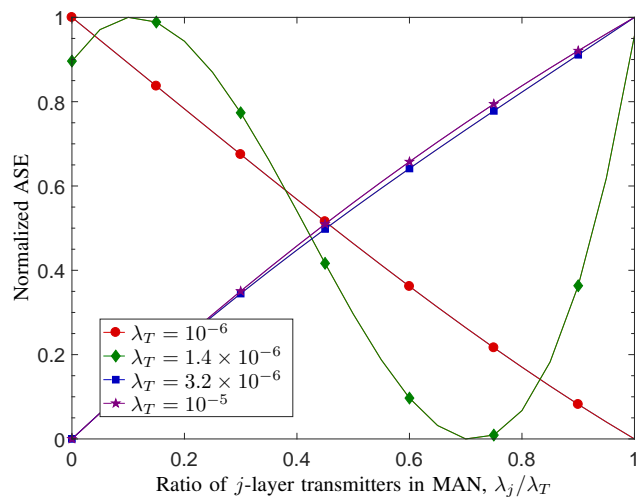}
			\vspace{-8mm}
		}
	\end{center}
\caption{
	Normalized \ac{ASE} of two layer \ac{MAN} according to the ratio of the $j$-layer transmitter density to the total density $\lambda\jtxn/\lambda_{T}$ with different total densities.
	$\lambda_T$ when 
	$h\jtxn=100$, $h\ktxn=200$, and $\tau=\oT\aS$.
}   \label{fig:T_Ratio}
\end{figure}

In this subsection, we show the \ac{STP} and the \ac{ASE} of the \ac{MAN} when the transmitter-oriented association and the ground receiver in the $0$-layer with density $\lambda_0=10^{-5}$ is considered.
We show the performance for a single layer \ac{AN} case in Figs.~\ref{fig:T_O_STP} and \ref{fig:T_O_ASE}, and then provide the performance for a two layer \ac{MAN} case in Figs.~\ref{fig:T_O_2Layer_ASE} and \ref{fig:T_Ratio}.

Figures \ref{fig:T_O_STP} and \ref{fig:T_O_ASE} show the \ac{STP} and \ac{ASE} of the single layer \ac{AN} as functions of the transmitter density $\lambda\jtxn$ and their altitude $h\jtxn$ when $\tau=\oR\aS$.
	The solid line with squars presents the optimal altitudes $h\jtxn^{*}$ that maximize \acp{STP} (in Fig.~\ref{fig:T_O_STP}) and the solid line with stars presents the optimal density $\lambda_j^{*}$ that maximizes \ac{ASE} (in Fig.~\ref{fig:T_O_ASE}) for different values of $\lambda\jtxn$.
	The dashed line in Fig.~\ref{fig:T_O_ASE} presents the upper bound of the optimal transmitter density, obtained from Corollary~\ref{cor:upper_bound}. Note that the same as the receiver-oriented association case, 
	the optimal transmitter density and its upper bound have the same trend, which decreases with the altitude of the \ac{AN}.
%
%
%
Furthermore, in Fig.~\ref{fig:T_O_ASE}, we can see that the optimal transmitter density in terms of the \ac{ASE} exists due to following reasons. 
	For small transmitter density $\lambda_j$, when $\lambda_j$ increases, 
	the impact of increasing number of the transmitting links in the network is large, 
	so ASE increases with $\lambda_j$. 
	However, for large $\lambda_j$, when $\lambda_j$ increases, 
	the impact of increasing interfering nodes and increasing their LoS probabilities to a receiver becomes 
	more critical than the increasing number of the transmitting links, 
	so the ASE decreases with $\lambda_j$.
Note that the optimal transmitter density in terms of the \ac{STP} is zero,
	since the larger transmitter density gives the more interfering nodes, 
	while the main link distance is not changed (as the transmitter-oriented association is used).

Figure~\ref{fig:T_O_2Layer_ASE} shows the \ac{ASE} the of two layer \ac{MAN} as functions of the $j$-layer transmitter density $\lambda\jtxn$ and the $k$-layer transmitter density $\lambda\ktxn$ when $h\jtxn=100$, $h\ktxn=200$, and $\tau=\oT\aS$.
The line marked with stars shows the optimal transmitter density of the $k$-layer $\lambda\ktxn^{*}$ for different values of $\lambda\jtxn$. 
	The lines marked with diamonds show the cases of having given values of the total transmitter density, i.e., 
	$\lambda\jtxn +\lambda\ktxn = \lambda_T$, 
	and the optimal density pairs, i.e., $(\lambda_j, \lambda_k)^*$ is marked with circles
	for different $\lambda_T$.
	From Fig. \ref{fig:T_O_2Layer_ASE}, we can see that 
	when $\lambda_T$ is large (e.g., $\lambda_T=10^{-5}$ and $\lambda_T=10^{-4.5}$), $\lambda\jtxn^{*}>0$ and  $\lambda\ktxn^*=0$. 
	However, when $\lambda_T$ is small (e.g., $\lambda_T=10^{-6}$),  
	$\lambda\ktxn^*>0$ and $\lambda\jtxn^{*}=0$.

In order to further clarify the relationship between the total transmitter density $\lambda_{T}$ and the optimal densities of each layers, 
	we present the normalized \ac{ASE} in Fig.~\ref{fig:T_Ratio}
	as a function of the ratio of the $j$-layer transmitter density to the total density $\lambda\jtxn/\lambda_{T}$ for different values of the total density $\lambda_{T}$.
	Here the normalized \ac{ASE}, $\mathcal{S}_{\lambda_T}^{\mathcal{N}}(\rho)$, is defined as
\begin{equation}
	\mathcal{S}_{\lambda_T}^{\mathcal{N}}(\rho)
	=
		\frac{\mathcal{S}_{\lambda_T}(\rho)-\min\limits_{\rho'}{\mathcal{S}_{\lambda_T}(\rho')}}
		{\max\limits_{\rho'}{\mathcal{S}_{\lambda_T}(\rho')}-\min\limits_{\rho'}{\mathcal{S}_{\lambda_T}(\rho')}}
\end{equation}
where $\mathcal{S}_{\lambda_T}(\rho)$ is the \ac{ASE} when the total density $\lambda_T$ and the ratio of $j$-layer transmitter density, $\rho=\lambda\jtxn/\lambda_T$, is given. 
Here, the normalized \ac{ASE} is a linear transform that makes the \ac{ASE} to have values between $[0,1]$, for the optimal ratio visualization. 
%
%
From Fig.~\ref{fig:T_Ratio}, we can see that 
	when the total transmitter density $\lambda_{T}$ is high, the optimal is to use the lower \ac{AN} only, i.e., $\lambda_j^{*}=\lambda_{T}$.
	On the other hand, when $\lambda_{T}$ is low, 
	the optimal is to use the higher \ac{AN} only, i.e., $\lambda_k^{*}=\lambda_{T}$.
	However, when $\lambda_{T}$ is neither high nor low such as $\lambda_{T}=1.4\times10^{-6}$, it is better to use the two layer \ac{MAN} instead of the single layer \ac{AN}, 
	which is the same as the \ac{STP} of the \ac{MAN} with the receiver-oriented association. 



	\section{Conclusion}
This paper establishes a foundation for the \ac{MAN} 
accounting for the different \ac{UAV} densities, altitudes, and transmission power in each layer \ac{AN}. 
After modeling the \ac{MAN} with the association rules and the channel, suitable for various scenario of the \ac{MAN},
we newly analyze the association probability, the main link distance distribution, and the Laplace transform of the interference. 
We then analyze the \ac{STP} and the \ac{ASE} of the \ac{MAN}, 
and also provide the upper bounds of the optimal \ac{UAV} densities that maximize the \ac{STP} and the \ac{ASE}, 
which is decreasing with the altitude of the \ac{AN} and determined independently without the effect of other layer \ac{UAV} densities. 
Finally, in the numerical results, we provide insights on the efficient design of the \ac{MAN}.
Specifically, we show that 
the optimal altitude of each \ac{AN}, maximizing the \ac{ASE}, decreases with the \ac{UAV} density,
and also the optimal \ac{UAV} density decreases with the altitude of the \ac{AN}. 
The optimal \ac{UAV} density of each \ac{AN}, maximizing the \ac{STP}, also decreases with the altitude of the \ac{AN}
for the receiver-oriented association case, while it becomes zero for the transmitter-oriented association case. 
We also show that when the total density of the \acp{UAV} is given, 
the optimal design of the \ac{MAN} is single \ac{AN} with the lower and the higher altitudes for large and small total densities, respectively, 
whereas the optimal design is to use the multiple layers when the density is neither large nor small.



	\begin{appendix}\label{app:und}
		\subsection{Proof of Corollary~\ref{cor:upper_bound}}\label{app:upper_bound}
		In order to derive the upper bound of the optimal densities, we get the derivatives of the \ac{STP} and the \ac{ASE} with respect to the transmitter density. 
		Then, we obtain the range of the densities that reduce the \ac{STP} and the \ac{ASE}, which gives the upper bound.  
		In this proof, we use following notation, which is not used in the rest of the paper.
		\begin{align}
			\mathcal{C}\ijtau\Co (y)=\AssoM p\ijtau\Co(y) f_{\Yoa}(y)
		\end{align}
		\subsubsection{receiver-oriented Association}
		From Lemma~\ref{L:r_Performance},
		when $o=\oR$, the derivative of the $k$-layer \ac{STP} with respect to the $j$-layer transmitter density is given by
		\begin{align}
		&\frac{\partial}{\partial \lambda_{j, \text{Tx}}} \STP{k}= 
	\sumAp \frac{\partial \fSi}{\partial \lambda_{j, \text{Tx}}} dy, \label{eq:A1}\\
		& \frac{\partial \fSi}{\partial \lambda_{j, \text{Tx}}}= 
		\begin{cases}
		  \frac{\fSi}{\lambda_{j, \text{Tx}}}\left(1-\lambda\jtx \pkj\right) & \text{for}\; i=j,\\
		  -\frac{\fSi}{\lambda_{j, \text{Tx}}}\pkj & \text{for}\; i\neq j,
		\end{cases} \label{eq:fSder} \\
		&\pkj =\label{eq:A3}\\
		&2\pi	\!\!	\sum_{\Tin} \left[ \int_{h_{ki}}^{\infty} x \rho_{ki}\To(x) dx  -\! \int_{\chi}^\infty\!\frac{x\rho_{ki}\To(x)}{1+ l_{ki}\Co (y) P_i x^{-\alpha\To}} dx \right] \nonumber
		\end{align}
		where, $\chi=\max\left(\chi_{j,i,\tau}^{(c,\cb)}(x),\;h_{ki}\right)$.
		When $i\neq j$, \eqref{eq:fSder} is always negative.
		Furthermore, when $i=j$, if the following inequality holds, \eqref{eq:fSder} is the negative.
		\begin{align}
		\underset{y, c}{\max} &\left[\frac{1}{ \pkjj    }\right] \leq \lambda_{j,\text{Tx}} \label{eq:L_bound} 
		\end{align}
		Here, $\pkjj$ increases with $y$ and $\phi_{kj,\tau}\coL(j,y)<\phi_{kj,\tau}\coN(j,y)$, therefore, $\pkjj$ has minimum at the $c=(\text{L})$ and $y=\hkj$, of which minimum is given by		
		\begin{align}
		\epsilon_{kj}=\phi_{kj,\tau}\coL (j,\hkj)  .\label{eq:A_bound_8}
		\end{align}
		Therefore, when $1/\epsilon_{kj}\leq\lambda\jtx$, the \ac{STP} of the $k$-layer is always decreased with the density of the transmitter in the $j$-layer, which gives the upper bound of the optimal density that maximizes the \ac{STP}. 
		In addition, the derivative of the \ac{ASE} is given by
		\begin{align}
				&\frac{\partial}{\partial \lambda_{j, \text{Tx}}} \ASE{k}= 
		\lambda_{k,\text{Rx}}\sumAp R_{ki} \frac{\partial \fSi}{\partial \lambda_{j, \text{Tx}}}  \label{eq:A5}.
		\end{align}
		The density of the receiver and the data rate are independent with the density of the transmitter. Therefore, if the inequality \eqref{eq:L_bound} holds, the \ac{ASE} decreases with the density of the transmitter, which gives the same upper bound of the density that maximizes the \ac{STP}. 
		
		\subsubsection{transmitter-oriented Association}		
When $o=\oT$, the derivative of the \ac{STP} of the $k$-layer with respect to the density of the $j$-layer transmitter is given by
\begin{align}
&\frac{\partial}{\partial \lambda_{j, \text{Tx}}} \STP{k}= 
\sumApj \frac{\partial \fSj}{\partial \lambda_{j, \text{Tx}}}  \label{eq:ddd}\\
& \frac{\partial \fSj}{\partial \lambda_{j, \text{Tx}}}= - \fSj \thkj \label{eq:fffff} \\
&\thkj =\label{eq:A4} \\
&2\pi		\sum_{\Tin}  \int_{\chi}^\infty x\rho_{ij}\To(x)\left(1-\frac{1}{1+ l_{ij}\Co (y) P_j x^{-\alpha\To}}\right) dx, \nonumber
\end{align}
where $\chi=\max (\chi_{k,j,\tau}^{(c,\cb)}, \hij)$. Therefore, the \ac{STP} always decreases with the transmitter density, which gives the optimal density as zero.
In addition, the derivative of the \ac{ASE} is given by
\begin{align}
&\frac{d}{d \lambda_{j, \text{Tx}}} \ASE{k}= 
\sumApj R_{ki} \frac{\partial }{\partial \lambda_{j, \text{Tx}}} \Big(\lambda_{k,\text{Tx}} \fSj\Big), \\
& \frac{\partial \Big(\lambda_{k,\text{Tx}} \fSj\Big)  }{\partial \lambda_{j, \text{Tx}}} =\nonumber \\
 &\begin{cases}
\fSj\left(1- \lambda\jtx\thkj \right)  
&\text{for}\; k=j,\\
- \lambda\ktx  \fSj  \thkj & \text{for}\;k\neq j.
\end{cases}
\end{align}
When $k\neq j$, the gradient is always negative which gives the optimal density as $0$.
When $k= j$, the range of the density that gives negative gradient is given by
\begin{align}
\underset{i, y, c}{\max} &\left[\frac{1}{ \thkj    }\right] \leq \lambda_{j,\text{Tx}}.\label{eq:A_bound_7} 
\end{align}
Note that we maximize over $i\in\mathcal{K}$. Here, as $\thkj$ increases with $y$ and $\theta_{kj,\tau}\coN(i,y)<\theta_{kj,\tau}\coL(i,y)$, the minimum is given as 
\begin{align}
\epsilon_{ki}=\theta_{kj,\tau}\coL(i,h_{ki})  .\label{eq:A_bound_8}
\end{align}
Therefore, we get the upper bound of the transmitter density that maximize the \ac{STP} and the \ac{ASE}.

		\subsection{Proof of Corollary~\ref{cor:dec}}\label{app:dec}
		From Corollary~\ref{cor:upper_bound}, $\epsilon_{ij}$ is given by
		\begin{align}
		&	\epsilon_{ij} = \int_{0}^\infty y  \left(1-q(y, \hij)\right) dy, \nonumber \\
		&q(y,\hij)=\frac{\varrho\ijcoL(y)}{1+\frac{\beta_{ij}\hij^{\alpha\coL}}{ (y^2+\hij^2)^{\alpha\coL/2}}}+ \frac{\varrho\ijcoN(y)}{1+\frac{\beta_{ij}\hij^{\alpha\coL}}{ (y^2+\hij^2)^{\alpha\coN/2}}}. \label{tmp:eqg}
		\end{align}
		We use the integration by substitution for $\epsilon_{ij}$ as $y^2=x^2-\hij^2$, hence,  the modified \ac{LoS} probability is $\varrho\ijcoL(y)=\rho\ijcoL\left(\sqrt{y^2+\hij^2}\right)$ which increase with $\hij$ for given $y$.
		Therefore, $\epsilon_{ij}$ increases with $\hij$ if $q(y, \hij)$ decreases with $\hij$ for all $y\in [0,\infty)$.
		The function $q(y, \hij)$ is further reformulated as
		\begin{align}
		&q(y, \hij)
		=\left(1+ \beta_{ij} \left(\frac{\hij^2}{\hij^2+y^2}\right)^{\alpha\coL/2}\right)^{-1} +\\
		&\frac{ \varrho\ijcoN(y) \beta_{ij} \hij^{\alpha\coL}\left(\left(\hij^2+y^2\right)^{\alpha\coN/2}-\left(\hij^2+y^2\right)^{\alpha\coL/2} \right)}
		{\left( \beta_{ij} \hij^{\alpha\coL} + \left(\hij^2+y^2\right) ^{\alpha\coN/2} \right) \left( \beta_{ij} \hij^{\alpha\coL}+ \left(\hij^2+y^2\right)^{\alpha\coL/2} \right)  } \nonumber
		\end{align}
		We use $\varrho\ijcoL (y)+\varrho\ijcoN(y)=1$.
		The upper part decreases with $\hij$ since $\frac{\hij^2}{\hij^2+y^2}$ increases with $\hij$. 
		Furthermore, the lower part decreases with $\hij$ if $1<\hij^2+y^2$ and $1<\beta_{ij}\hij^{\alpha\coL}$. Therefore, $\epsilon_{ij}$ decreases with $\hij$ if $1<\hij$ and $1<\beta_{ij}\hij^{\alpha\coL}$.

	\end{appendix}

	\bibliographystyle{IEEEtran}

	\bibliography{./StringDefinitions,./IEEEabrv,./myBibDongsun}

\end{document}